\documentclass[a4paper]{article}

\usepackage{arxiv}

\usepackage[authoryear,longnamesfirst]{natbib}
\usepackage{hyperref}
\usepackage{subfigure}
\usepackage[ruled,vlined]{algorithm2e}
\usepackage{amssymb}
\usepackage{amsmath} 
\usepackage{booktabs}       
\usepackage{amsfonts}       
\usepackage{graphicx}
\usepackage{tabularx}

\title{Bayesian estimation and reconstruction of marine surface contaminant dispersion}

\author{
    Yang Liu  \\
	Dept. of Aeronautical and Automotive Engineering\\
	Loughborough University\\
	Leicestershire, LE11 3TU \\
	\texttt{Y.Liu6@lboro.ac.uk} \\
	\And
	Christopher M.~Harvey \\
	Dept. of Aeronautical and Automotive Engineering\\
	Loughborough University\\
	Leicestershire, LE11 3TU \\
	\texttt{c.m.harvey@lboro.ac.uk} \\
	\And
	Frederick E.~Hamlyn \\
	Dept. of Aeronautical and Automotive Engineering\\
	Loughborough University\\
	Leicestershire, LE11 3TU \\
	\texttt{f.e.hamlyn@lboro.ac.uk} \\
    \And
	Cunjia Liu \thanks{Corresponding author}\\
	Dept. of Aeronautical and Automotive Engineering\\
	Loughborough University\\
	Leicestershire, LE11 3TU \\
	\texttt{c.liu5@lboro.ac.uk} \\
}

\begin{document}
\maketitle

\begin{abstract}
Discharge of hazardous substances into the marine environment poses a substantial risk to both public health and the ecosystem. In such incidents, it is imperative to accurately estimate the release strength of the source and reconstruct the spatio-temporal dispersion of the substances based on the collected measurements. In this study, we propose an integrated estimation framework to tackle this challenge, which can be used in conjunction with a sensor network or a mobile sensor for environment monitoring. We employ the fundamental convection-diffusion partial differential equation (PDE) to represent the general dispersion of a physical quantity in a non-uniform flow field. The PDE model is spatially discretised into a linear state-space model using the dynamic transient finite-element method (FEM) so that the characterisation of time-varying dispersion can be cast into the problem of inferring the model states from sensor measurements. We also consider imperfect sensing phenomena, including miss-detection and signal quantisation, which are frequently encountered when using a sensor network. This complicated sensor process introduces nonlinearity into the Bayesian estimation process. A Rao-Blackwellised particle filter (RBPF) is designed to provide an effective solution by exploiting the linear structure of the state-space model, whereas the nonlinearity of the measurement model can be handled by Monte Carlo approximation with particles. The proposed framework is validated using a simulated oil spill incident in the Baltic sea with real ocean flow data. The results show the efficacy of the developed spatio-temporal dispersion model and estimation schemes in the presence of imperfect measurements. Moreover, the parameter selection process is discussed, along with some comparison studies to illustrate the advantages of the proposed algorithm over existing methods.
\end{abstract}


\section{Introduction}\label{sec1}
A wide spectrum of hazardous substances, ranging from chemicals and microplastics to heavy metals and radioactive materials, can be found in the marine environment, causing significant damage to public health and the ecosystem. Considerable research effort has therefore been dedicated to understanding the composition and impact of water pollution, to tracing pollution sources, and to characterising the extent of pollutant dispersion. For example, the risk of hazardous heavy metals (such as Pb, Cu, Zn, and Hg) was assessed through samples from sea surface sediments, and subsequently the influencing factors were identified \citep{el2023assessment,hu2023surface}. Microplastics in the marine environment have raised many concerns \citep{chen2023global} because of their high abundance and persistent degradation, and so understanding the accumulation, dispersion, and transportation of microplastics has become increasingly important \citep{schwarzplastics}. This also applies to other types of pollutants discharged into water environments, where identifying the emission source and modelling the pollution dispersion is required to inform and enable more effective countermeasures and regulatory actions. Recent examples of research in this direction include the estimation of pesticides emissions into river systems \citep{ZHANG2020113660} and mapping the spatio-temporal distribution of antimicrobial resistant organisms in different aquatic environments \citep{ASADUZZAMAN2022154890}. The dispersion of multiple contaminants directly released into water has also been formulated along shipping lanes and in harbors \citep{maljutenko2021modelling}.

Environment monitoring often relies on point measurement data collected either through manual sampling or in-situ sensor networks to infer the environment phenomena. One common approach is spatial interpolation between measurement points, whereby missing values are estimated using techniques such as Kriging \citep{OLIVER1990}, Gaussian process \citep{rasmussen2006gaussian}, and more recently, machine learning algorithms (e.g., \citealp{LI2023162336}). When it comes to characterising the emission source of hazardous substances, a more effective approach is to formulate a source term estimation (STE) task. STE strategies can be roughly categorised as either optimisation-based or filter-based, both of which have found many successful applications, including in atmospheric and water environments \citep{LEWIS2021144029,hutchinson2018information,ZHU2021117497,anshuman2023parallel,WANG2018759}. It is noted that the majority of STE studies focus on estimating the position and release strength of the source, as in a static parameter estimation problem. In practice, sometimes a spatio-temporal dispersion field needs to be reconstructed in the region of interest (ROI), in addition to the source parameters, so that a more comprehensive understanding of the pollution dynamics can be built over the whole ROI and more information can be provided for the subsequent actions such as pollutant cleanup (e.g., see optimizing oil spill cleanup resources in \citealp{grubesic2017optimizing}).

To accurately estimate the source term and reconstruct the dispersion process, a mathematical model is required to represent the spatio-temporal field. The general convection-diffusion partial differential equation (PDE) that models the dispersion of a physical quantity in a flow field has been widely employed to capture the spatio-temporal evolution of pollutants. For example, the dispersion of nitrogen in the Great Barrier Reef has been described with a PDE; and the finite-element method (FEM) with deep neural networks has been employed to predict spatial maps of nitrogen at the forecast time steps \citep{jahanbakht2022nitrogen}. In \cite{hodgson2022model}, a PDE has been adopted to model the fluid velocity, and the dynamics of oil particle position has been formulated accordingly.

The analytical solution of a PDE is usually unavailable, which poses extra challenges for STE and dispersion reconstruction problems: To perform Bayesian inference of the source parameters or reconstruct the dispersion from the set of point measurements, a solvable model, which can incorporate the stochastic nature of PDE parameters and states, needs to be established. The FEM is specifically suitable in the addressed spatio-temporal estimation problem. Using the FEM, an adaptive Bayesian reference method has been provided in \cite{ALBANI2021118039} to characterise the source release, in which the normalised $l_2$ discrepancy between actual and the calculated measurement concentrations has been employed in the Metropolis-Hastings algorithm, and the state-dependent scaling method has been used to accelerate the convergence process of the algorithm. The FEM has also been used in \cite{forti2022unknown} to approximate the solution to the PDE, with the ROI partitioned into mesh elements and the system state constructed from the values of the field at the mesh nodes. Furthermore, the classical Kalman filter has been selected to estimate the source terms, and the source identifiability has been analysed by resorting to the transfer functions. It is noted that the measurement outputs in \cite{forti2022unknown,ALBANI2021118039} can be modelled as a linear combination of the system states with additive Gaussian noises, which permits the direct application of the classical Kalman filter. Unfortunately, in practice, measurements are often imperfect, especially when considering a sensor network with connectivity limitations, and this needs to be systematically considered in the filter development.

In this paper, we aim to design a Bayesian estimation framework for estimating and reconstructing the spatio-temporal dispersion of a point release source, which can be used in conjunction with a sensor network or a mobile sensor for environment monitoring. Therefore, some practical issues associated with sensing and data transmission need to be considered. One phenomenon of wireless signal transmission is quantisation. In reality, the precision of the transmitted data is inherently limited by the bandwidth, which necessitates the use of quantisers during analog-to-digital conversion in signal transmission. The finite word length of quantised signals naturally leads to some information loss. To handle the quantisation-induced loss, many estimation strategies have been developed, see, for example, \cite{liu2022novel,li2021distributed} and the references therein. In environment monitoring studies, the quantisation issue has also garnered some research attention: In the experimental investigation on the effects of polyethylene microplastics on mammals \citep{sun2021effects}, the measurement of colon mucin density has been quantised with the Alcian blue periodic acid Schiff. The limits of quantisation (LOQ) have been analysed in concentration measurement of the Steroid 11-Desoxycorticosterone drug residues in water for trace determination \citep{alsaggaf2022electrooxidation}. The LOQ have also been studied in the vapor phase mercury emissions of combustion sources by using isotope dilution inductively coupled plasma mass spectrometry \citep{long2020provision}. Spatio-temporal modelling and STE problems subject to signal quantisation have not drawn enough research attention yet, and there is much room for further improvement.

Miss detection, also referred to as missing measurement, is a another common sensing phenomenon in practice. It may arise from intermittent sensor failures, transmission delay, communication loss, and other factors like local turbulence in atmospheric sensing \citep{hutchinson2018information}. So far, much research effort has been devoted to the analysis and synthesis of systems with miss detection \citep{dong2020fault,gao2022survey}. In the field of environmental monitoring, miss detection occurring in sensors has received some attention: In \cite{govindaraju2020monitoring}, the missing detection rate has been analysed with respect to a wireless sensor network consisting of LEDs, temperature, lightning and humidity sensors, which have been deployed on the surface of high voltage insulators to monitor the state of pollution. The miss detection phenomenon has been discussed in \cite{amin2021viral} for bio-sensors, which can measure diffused virus-laden aerosol concentration, where the partial absorption and reflection at each side of the room has been considered. By using a multiscale multidimensional residual convolutional neural network, the effects of miss detection for oil spills has been reduced with optical remote sensing imagery in \cite{seydi2021oil}. To mitigate the effects of the miss detection, some data-driven interpolation methods have been employed in pollution monitoring experiments, for example, a temporal correction procedure has been developed for indoor radon concentrations \citep{ivanova2019modelling}; and different approaches have been compared in imputing the missing daily ground-level particulate matter concentration data in air quality sensing \citep{arowosegbe2021comparing}. The introduction of miss detection significantly complicates the estimator design in terms of both the structure and the algorithms, and this constitutes another motivation of the current investigation.

To effectively address the challenges induced by signal quantisation and miss detection, it is crucial to employ appropriate estimation techniques, such as particle filtering or sequential Monte Carlo simulation, which can account for the involved stochasticity and nonlinearity. In conventional particle filtering techniques, high computational cost is one of the main disadvantages, especially in the high-dimensional model obtained with the FEM. As a representative Bayesian estimation method, the Rao-Blackwellised particle filter (RBPF) \cite{sarkka2007rao} can achieve less estimation error variance than conventional Bayesian estimator with pure Monte Carlo sampling \cite{casella1996rao}. The main idea of RBPF is to employ the Rao-Blackwell theorem in the sampling process for a particle filter by marginalising out some variables, and thus decrease the need for a large number of particles. In the RBPF proposed in this paper, the effects of miss detection and quantisation can be incorporated in the calculation of particle weights by adjusting the likelihood function of the measurement with respect to each particle, facilitating the STE and the reconstruction of the spatio-temporal dispersion processes.

Motivated by the discussions above, in this paper, the RBPF is developed to estimate the source term and reconstruct the dispersion distribution in the presence of miss detection and quantisation. The dynamic transient FEM with Galerkin planar triangle elements is used to solve the PDE of the dispersion process, taking into account the non-uniform flow field of the marine surface environment. A Bernoulli sequence is employed to represent stochastic miss measurement. A uniform quantiser is associated with each sensor node to reduce the transmission burden. An RBPF is constructed by incorporating the measurements from all the sensor nodes, where the weight of each particle is determined by resorting to the characteristics of the miss detection and the quantisation parameters. To showcase the effectiveness of the proposed approach, an oil spill example is presented based on real current data from the Baltic sea. The main contributions of this paper are highlighted as follows: 1) a detailed dynamic transient FEM model with Galerkin planar triangle elements is provided for the convection-diffusion dispersion PDE; 2) miss detection and quantisation are newly considered in the estimation problem; and 3) an efficient RBPF algorithm is proposed to simultaneously address the STE and dispersion reconstruction problems.

\section{Methods} \label{sec2}

In this section, a dynamic dispersion model is first established by solving the general convection-diffusion PDE using the dynamic transient FEM with Galerkin planar triangle elements. Then, a measurement model is proposed, incorporating miss detection and signal quantisation. Finally, the RBPF is designed to solve the estimation and dispersion field reconstruction problem.

\subsection{Dispersion model} \label{subsec2-2}
Consider the convection-diffusion PDE, given by
\begin{align}\label{PDE1}
&\dot{c}(p,t) - \lambda\nabla^2 c(p,t) + \nabla \cdot (\mathbf{v}(p,t) \, c(p,t)) = f(p,t)
\end{align}
with the initial condition
\begin{align}\label{PDE2}
c(p,0)=c_0(p)
\end{align}
where $c(p,t)$ is the scalar field of the environmental variable of interest (e.g., pollutant concentration) defined over the space-time domain $\Omega\times R$. The space domain $\Omega$ is assumed to have a smooth boundary $\partial\Omega$, and $p\in\Omega$ is the $d$-dimensional position vector $(d\in\{1,2,3\})$. Time is denoted by $t\in R$. The symbols $\nabla\cdot$ and $\nabla^2$ represent the  divergence and Laplacian operators, respectively, and the $\dot{c}$ denotes the time derivative, $\partial c/\partial t$. The $\lambda$ diffusion coefficient is constant in time and spatially uniform; the advection velocity vector $\mathbf{v}(p,t)$ may vary in time and spatially. The point source input $f(p,t)$ is modelled as
\begin{align}\label{PDE4}
f(p,t)=u(t) \, \delta(p-p_s(t))
\end{align}
with unknown intensity $u(t)$ and position $p_s(t)\in\Omega$, where $\delta(p-p_s(t))$ is the Dirac delta function centred on $p_s(t)$, and so $f(p,t)$ is zero everywhere except at $p_s$.

Since an analytical solution of Eq.~\eqref{PDE1} is unachievable in most cases, an approximate solution is sought, and so the FEM is adopted. Linear planar triangular elements (with three nodes) are adopted in this paper. Therefore, within a given element, assume an approximate solution for $c$ of the form
\begin{align}\label{source}
c(p,t)=\sum\limits_{i=1}^3 c_i(t)\mathcal{B}_i(p)
\end{align}
where $c_i(t)$ is $i$th nodal value of $c$, which is to be estimated, and $\mathcal{B}_i(p)$ represents the element's shape functions.

If the nodes are located at $(x_i,y_i)$ with $i=1,2,3$, then the shape functions are
\begin{align}
\begin{aligned}\label{triA}
\mathcal{B}_1(p)=&\frac{1}{2S}\left[-y_{32}(x-x_2)+x_{32}(y-y_2)\right]\\
\mathcal{B}_2(p)=&\frac{1}{2S}\left[y_{31}(x-x_3)-x_{31}(y-y_3)\right]\\
\mathcal{B}_3(p)=&\frac{1}{2S}\left[-y_{21}(x-x_1)+x_{21}(y-y_1)\right]
\end{aligned}
\end{align}
where $x_{ij}=x_i-x_j$, $y_{ij}=y_i-y_j$, and
\begin{align}
S=\left|x_{21}y_{32}-x_{32}y_{21}\right|/2
\end{align}
is the element area. Higher element densities can characterise the ROI more accurately at the cost of heavier computational burden due to the higher problem dimensionality.

Now let us employ the Galerkin method to derive the element stiffness and mass matrices. For a planar linear triangular element, the Galerkin formulation of Eq.~\eqref{PDE1} is ($i=1,2,3$)
\begin{align}\label{Galerkin1}
\begin{split}
&\iint_S \left[\dot{c}(p,t) - \lambda \left( \frac{\partial^2 c}{\partial x^2} + \frac{\partial^2 c}{\partial y^2} \right) \right. \\
&\left. {} + \left(u \frac{\partial c}{\partial x} + v \frac{\partial c}{\partial y}\right) - f(p,t) \right] \mathcal{B}_i(p) \, dx \, dy = 0.
\end{split}
\end{align}
In this work, the advection velocity $\mathbf{v}$ has been taken as constant $u\,\mathbf{i} + v\,\mathbf{j}$ within each element. Higher order approximations can be introduced for the velocity field for greater accuracy in the case of strong advection velocity gradients.

Linear elements are not suitable to approximate the second-order partial derivatives of the diffusion term in Eq.~\eqref{Galerkin1} and so Green's Theorem in a plane,
\begin{align}\label{green}
\iint_S\left(\frac{\partial G}{\partial x}-\frac{\partial F}{\partial y}\right)dxdy=\oint_\Gamma\left(Fdx+Gdy\right),
\end{align}
is used to modify the Galerkin formulation and reduce the order of the second-order derivative terms. Note that $\Gamma$ in Eq.~\eqref{green} represents the element boundary. Choosing
\begin{align}
F=-\displaystyle\frac{\partial c(p,t)}{\partial y}\mathcal{B}_i(p)\quad\text{and}\quad G=\displaystyle\frac{\partial c(p,t)}{\partial x}\mathcal{B}_i(p)
\end{align}
provides, after some algebraic manipulation,
\begin{align}
\begin{split}
&\iint_S \mathcal{B}_i(p) \left(\frac{\partial^2 c(p,t)}{\partial x^2}+\frac{\partial^2 c(p,t)}{\partial y^2}\right) \, dx \, dy \\
&{} =\oint_\Gamma \mathcal{B}_i(p) \left(\frac{\partial c(p,t)}{\partial x} dy - \frac{\partial c(p,t)}{\partial y} dx\right) \\
&{} -\iint_S\left(\frac{\partial c(p,t)}{\partial x}\frac{\partial \mathcal{B}_i(p)}{\partial x}+\frac{\partial c(p,t)}{\partial y}\frac{\partial \mathcal{B}_i(p)}{\partial y}\right) \, dx \, dy,
\end{split}
\end{align}
which after being introduced into Eq.~\eqref{Galerkin1} gives the modified Galerkin formulation of Eq.~\eqref{PDE1} as
\begin{align}\label{Green6}
\begin{split}
&\iint_S \mathcal{B}_i(p) \, \dot{c}(p,t) \, dx \, dy \\
&{} + \lambda \iint_S \left(\frac{\partial c(p,t)}{\partial x}\frac{\partial \mathcal{B}_i(p)}{\partial x} + \frac{\partial c(p,t)}{\partial y}\frac{\partial \mathcal{B}_i(p)}{\partial y}\right) \, dx \, dy \\
&{} + \iint_S \mathcal{B}_i(p) \left(u \frac{\partial c(p,t)}{\partial x} + v \frac{\partial c(p,t)}{\partial y} \right) \, dx \, dy \\
&{} = \iint_S \mathcal{B}_i(p) f(p,t) \, dx \, dy \\
&{} + \lambda \oint_\Gamma \mathcal{B}_i(p) \left(\frac{\partial c(p,t)}{\partial x} dy - \frac{\partial c(p,t)}{\partial y} dx\right).
\end{split}
\end{align}
Now, the second-order derivatives in Eq.~\eqref{Galerkin1} have been replaced by the product of first-order derivatives in Eq.~\eqref{Green6}, and so linear elements can be suitably employed. Introducing Eq.~\eqref{source} into Eq.~\eqref{Green6}, and defining $$\Bar{c}(t)=\begin{bmatrix} c_1(t) & c_2(t) & c_3(t) \end{bmatrix}^\text{T}$$ allows Eq.~\eqref{Green6} to be rewritten in matrix form as
\begin{align}\label{Galerkin8}
\mathcal{A}\dot{\bar{c}}(t)+\mathcal{D}\bar{c}(t)
=\iota(t)+d(t)
\end{align}
where the elements of $\mathcal{A}$, $\mathcal{D}$, $\iota(t)$ and $d(t)$ are obtained by cycling $i=1,2,3$ and $j=1,2,3$ in the following:
\begin{gather}
\mathcal{A} = \iint_S \mathcal{B}_i(p) \, \mathcal{B}_j(p) \, dx \, dy, \label{mass_matrix}\\
\begin{split}
\mathcal{D}=
&\lambda \iint_S \left(\frac{\partial \mathcal{B}_j(p)}{\partial x}\frac{\partial \mathcal{B}_i(p)}{\partial x} + \frac{\partial \mathcal{B}_j(p)}{\partial y}\frac{\partial \mathcal{B}_i(p)}{\partial y}\right) \, dx \, dy \label{stiffness_matrix}\\
&{} + \iint_S \mathcal{B}_i(p) \left(u \frac{\partial \mathcal{B}_j(p)}{\partial x} + v \frac{\partial \mathcal{B}_j(p)}{\partial y} \right) \, dx \, dy
\end{split} \\
\iota(t) = \iint_S \mathcal{B}_i(p) f(p,t) \, dx \, dy \label{force_vector}\\
d(t) = \lambda \oint_\Gamma \mathcal{B}_i(p) \left(\frac{\partial c(p,t)}{\partial x} dy - \frac{\partial c(p,t)}{\partial y} dx\right) \label{bcs}
\end{gather}
$\mathcal{A}$, $\mathcal{D}$, $\iota(t)$ and $d(t)$ are commonly called the mass matrix, stiffness matrix, force vector and boundary conditions vector, respectively. Based on Eq.~\eqref{triA}, full integration of Eqs.~\eqref{mass_matrix}--\eqref{force_vector} gives
\begin{gather}
\mathcal{A}=\frac{S}{12}\left[\begin{array}{ccc} 2 & 1 & 1\\ 1 & 2 & 1\\ 1 & 1 & 2\end{array}\right] \label{mass_matrix2}\\
\begin{split}
\mathcal{D}=&\frac{1}{6} \begin{bmatrix}
vx_{32}-uy_{32} & uy_{31}-vx_{31} & vx_{21}-uy_{21}\\ vx_{32}-uy_{32} & uy_{31}-vx_{31} & vx_{21}-uy_{21}\\ vx_{32}-uy_{32} & uy_{31}-vx_{31} & vx_{21}-uy_{21}
\end{bmatrix}\\
&{} +\frac{\lambda}{4S} \begin{bmatrix}
y_{32}^2 & -y_{32}y_{31} & y_{32}y_{21} \\
-y_{32}y_{31} & y_{31}^2 & -y_{31}y_{21} \\
y_{32}y_{21} & -y_{31}y_{21} & y_{21}^2
\end{bmatrix} \\
&{} +\frac{\lambda}{4S} \begin{bmatrix}
x_{32}^2 & -x_{32}x_{31} & x_{32}x_{21} \\
-x_{32}x_{31} & x_{31}^2 & -x_{31}x_{21} \\
x_{32}x_{21} & -x_{31}x_{21} & x_{21}^2
\end{bmatrix}
\end{split} \label{stiffness_matrix2}
\end{gather}
and, for the element containing the source input $f(p,t)$,
\begin{align}\label{force_vector2}
\iota(t) = \frac{S u(t)}{3}\begin{Bmatrix}
1 & 1 & 1
\end{Bmatrix}^\text{T},
\end{align}
which assumes the source input is uniformly $u(t)$ within the element. For all other elements that do not contain the source,
\begin{align}\label{force_vector3}
\iota(t) = \begin{Bmatrix}0 & 0 & 0\end{Bmatrix}^\text{T}.
\end{align}

Now that the linear planar triangle element has been established for the convection-diffusion PDE, Eq.~\eqref{Galerkin8} can be assembled across all the elements in the domain into
\begin{align}\label{assembled}
M\Dot{\hat{c}}(t)+N\hat{c}(t)=Q(p_s(t))u(t)+\hat{d}(t),
\end{align}
where
\begin{align}
\hat{c}(t) = \begin{bmatrix}
c_1(t) & c_2(t) & \ldots & c_C(t)
\end{bmatrix}^\text{T}
\end{align}
with $C$ denoting the total number of nodes, and where $M$, $N$, $Q(p_s(t))$ and $\hat{d}(t)$ are constructed by assembling $\mathcal{A}$, $\mathcal{D}$, $\iota(t)$ and $d(t)$, respectively, across every element. Far from any pollutant concentration or on a solid boundary (e.g., a shoreline), the spatial gradients, $\partial c/\partial x$ and $\partial c/\partial y$, are approximately zero, and so the $\hat{d}(t)$ boundary conditions can be taken as zero as long as either condition holds.

Forward Euler/explicit time integration can be used to write Eq.~\eqref{assembled} at each sampling time step. For convenience, a uniform sampling interval is used for both the dispersion and sensor models, that is, $t_k=k\delta t$ where $\delta t$ is the constant interval. The following discrete-time linear system can be established:
\begin{align}\label{dis}
M\left(\frac{\hat{c}_{k+1}-\hat{c}_k}{\delta t}\right) + N\hat{c}_{k} = Q(p_{s,k}) u_k. 
\end{align}
It follows naturally that
\begin{align}\label{ck+1}
\hat{c}_{k+1}=A\hat{c}_{k}+B(p_{s,k})u_k+w_k
\end{align}
where
\begin{align}\label{para1}
\begin{aligned}
A&=M^{-1}\left(M-\delta t N\right)\\
B(p_{s,k})&=\delta t M^{-1}Q(p_{s,k})
\end{aligned}
\end{align}
and $w_k$ is a process disturbance accounting for the errors induced by the FEM approximation. It is assumed that $w_k\sim \mathcal{N}(0,W_k)$ where $W_k>0$ is an available covariance matrix.

In this work, constant source strength $u_k$ is adopted; however, the actual value of the strength is unavailable to us. To compensate such a shortage, the source strength is assumed to satisfy
\begin{align}\label{upre}
u_{k}=u_{k-1}+\varpi_{k},
\end{align}
where $\varpi_{k}\sim \mathcal{N}(0,\Pi_{k})$ and $\Pi_{k}>0$ is a pre-defined covariance matrix. Stochastic noise is induced to cater for the unknown strength value.

For the system in Eq.~\eqref{ck+1}, an augmented system state is defined as $x_k\triangleq\left[\hat{c}_{k}^\text{T},u^\text{T}_{k}\right]^\text{T}$. Then, based on Eq.~\eqref{upre},
\begin{align}\label{xk+1}
x_{k+1}=\bar{A}(p_{s,k})x_{k}+\bar{w}_{k}
\end{align}
where
\begin{align}
\begin{aligned}
\bar{A}(p_{s,k})&\triangleq\left[\begin{array}{cc}A&B(p_{s,k})\\0&I\end{array}\right]\\
\bar{w}_{k}&\triangleq\left[w_{k}^\text{T},\varpi_{k}^\text{T}\right]^\text{T}.
\end{aligned}
\end{align}
Based on the definitions above, it is straightforward to see that
\begin{align}\label{covw}
\bar{W}_{k}\triangleq {\mathbb E}\left\{\bar{w}_{k}\bar{w}_{k}^\text{T}\right\}=\text{diag}\left\{W_k,\Pi_{k}\right\}.
\end{align}

Since the state $x_k$ consists of both nodal concentration and source strength, we can estimate $x_k$ well at each time step, and then the STE and dispersion field reconstruction problems can be solved simultaneously.

Finally, it is important to comment on the numerical stability of Eq.~\eqref{ck+1}. For explicit time integration, the Courant condition requires that
\begin{align}\label{stab_conv}
\delta t \le \frac{h}{\left|u\right|}
\end{align}
where $h$ is the element size and $u$ is the wave speed, or for a diffusion process, using the concept of diffusion time,
\begin{align}\label{stab_diff}
\delta t \le \frac{h^2}{2 \lambda}.
\end{align}
In general, the region of stability is defined by \cite{ascher1997} as
\begin{align}
\left| 1 - \Lambda \, \delta t \right| \le 1
\end{align}
which must hold for all the problem's $\Lambda$ eigenvalues. Since all the eigenvalues are positive for this problem, the condition becomes
\begin{align}\label{stab_gen}
\delta t \le \frac{2}{\Lambda_\text{max}}
\end{align}
where $\Lambda_\text{max}$ is the maximum eigenvalue of $\mathcal{A}^{-1} \mathcal{D}$. Eqs.~\eqref{stab_conv} and \eqref{stab_diff} are special cases of the general condition in Eq.~\eqref{stab_gen}.

The time step used may be less than that given by Eq.~\eqref{stab_gen}, but there exists a lower bound below which the solution exhibits spatial oscillations. More details are given in \cite{cui2016}. A reliable solution to avoid the instability, however, is to use the lumped mass matrix instead of the consistent mass matrix given in Eq.~\eqref{mass_matrix2}, that is,
\begin{align}
\mathcal{A}=\frac{S}{3}\left[\begin{array}{ccc} 1 & 0 & 0\\ 0 & 1 & 0\\ 0 & 0 & 1\end{array}\right] \label{mass_matrix3}.
\end{align}

Another instability is in the form of spurious spatial oscillations, which may occur when the P\'eclet number is larger than one. (The P\'eclet number is the ratio of inertial to diffusive forces.) For stability,
\begin{align}
    \text{Pe} \le 1 \quad\text{where}\quad \text{Pe} = \frac{\left| \mathbf{v} \right| h}{2\lambda}.
\end{align}
If convection is present, there is an element size above which the spurious spatial oscillations occur. This means that the instability can be solved by refining the mesh. Another solution is to artificially increase the diffusivity to give $\text{Pe}\le1$. Note that both solutions also decrease the critical time step by the Courant condition. Let the additional artificial diffusivity be $\lambda^\ast$ and then \citep{onate2000},
\begin{align}
    \lambda^\ast = \frac{\alpha \left| \mathbf{v} \right| h}{2} \quad\text{with}\quad \alpha\ge 1 - \frac{1}{\text{Pe}}.
\end{align}
It is easily confirmed the the Pe number is decreased to less than or equal to 1 by increasing $\lambda$ by $\lambda^\ast$.

\subsection{Measurement model} \label{subsec2-3}

For the dispersion process described by Eq.~\eqref{PDE1}, assume that there are $N$ sensors located at $s_j\in\Omega$ $(j=1,2,\ldots,N)$, and the discrete measurement can be formulated as
\begin{align}\label{PDE5}
y_{j,k}=h(c(s_j,t_k))+v_{j,k},
\end{align}
where time index $k\in\left\{1,2,\ldots\right\}$ such that $0<t_1<t_2<\cdots$, and $v_{j,k}$ is the measurement noise with known distribution. In this paper, we consider a 2-dimensional region, where the sensors are assumed to be static, i.e., $s_j$ is time-invariant, and the sensor noise follows $v_{j,k}\sim \mathcal{N}(0,V_{j,k})$, where the covariance matrix $V_{j,k}$ is available. For simplicity, we assume the sensors directly measure the concentration value at its location, i.e.,
\begin{align}\label{PDEadd}
y_{j,k}=c(s_j,t_k)+v_{j,k}.
\end{align}

Based on Eq.~\eqref{source}, $c(s_j,t_k)$ can be written
\begin{align}\label{measure}
c(s_j,t_k)=\sum\limits_{i=1}^C c_i(t_k)\mathcal{B}_i(s_j).
\end{align}

With the discretised model in Eq.~\eqref{dis}, the measurement can be then formulated as
\begin{align}\label{yk}
y_{j,k}=H_j \hat{c}_{k}+v_{j,k}.
\end{align}
where the matrix can be calculated based on the position of the sensor and the vertex index. For the augmented system state $x_k$, it follows naturally that:
\begin{align}\label{linearyk}
y_{k}=Hx_{k}+v_{k},
\end{align}
where
\begin{align*}
&y_k\triangleq\left[y^\text{T}_{1,k},\ldots,y^\text{T}_{N,k}\right]^\text{T},v_k\triangleq\left[v^\text{T}_{1,k},\ldots,v^\text{T}_{N,k}\right]^\text{T}, \\
& H\triangleq\left[\bar{H}^\text{T}_{1},\ldots,\bar{H}^\text{T}_{N}\right]^\text{T}, \bar{H}_j\triangleq\left[H_j,0\right].
\end{align*}
It is straightforward to see that the covariance of $v_k$ is $\bar{V}_k=\text{diag}\left\{V_{1,k},\ldots,V_{N,k}\right\}$.

The measurement model in Eq.~\eqref{linearyk} is linear with respect to the augmented state and the additive noise. To better reflect the imperfect measurements in practical sensor networks, the miss detection and uniform quantisation are considered. The sensor measurement subject to miss detection is described as \cite{gao2022survey}
\begin{align}\label{missyk}
y_{j,k}=\alpha_{j,k}H_j \hat{c}_{k}+v_{j,k},
\end{align}
where $\alpha_{j,k}$ is a stochastic variable which accounts for the miss detection. In our work, $\alpha_{j,k}$ is independent of all the other noises and obeys the following Bernoulli distribution
\begin{align}\label{alpha}\setlength{\arraycolsep}{0.2em}\left\{\begin{array}{rl}
{\rm Prob}\left(\alpha_{j,k}=1\right)=&\bar{\alpha}_{j,k},\\
{\rm Prob}\left(\alpha_{j,k}=0\right)=&1-\bar{\alpha}_{j,k},\end{array}\right.
\end{align}
with known scalar $\bar{\alpha}_{j,k}\in[0,1]$. Normally, we can consider a constant miss detection rate, such that $\bar{\alpha}_{0}=1-\bar{\alpha}_{j,k}$ and $\bar{\alpha}_{1}=\bar{\alpha}_{j,k}$.

To model the quantisation of the measurement $y_{j,k}$ in Eq.~\eqref{missyk} during data transmission, for a given scaling parameter $\eta_j>0$, the quantisation range is defined as 
\begin{align}\label{region}
{\mathcal R}=\left\{y_{j,k}\in{\mathbb R}:\left|y_{j,k}\right|\leq \eta_j \right\},
\end{align}
and for a positive scalar $\zeta_j$, the quantisation function $q_j(\cdot):{\mathbb R}\rightarrow{\mathbb R}$ is set as \cite{li2021distributed}
\begin{equation}
\label{quanti1}
q_j\left(y_{j,k}\right)=\left\{\begin{array}{ll}
\rho_{j,h} & \mathrm{if} \, \rho_{j,h}-\frac{\eta_j}{\zeta_j} \leq y_{j,k}<\rho_{j,h}+\frac{\eta_j}{\zeta_j}\\
\rho_{j,\zeta_j-1} & \mathrm{if} \, y_{j,k}=\eta_j
\end{array}
\right.
\end{equation}
where the quantisation levels are provided as
\begin{align}\label{quanti2}
{\mathcal L}_j=\left\{\rho_{j,h}: \rho_{j,h}=-\eta_j+\frac{(2h+1)\eta_j}{\zeta_j},h=0,\ldots,\zeta_j-1 \right\}.
\end{align}
In this case, the signal received at the filter can be denoted as $\hat{y}_k\triangleq\left[\hat{y}^\text{T}_{1,k},\ldots,\hat{y}^\text{T}_{N,k}\right]^\text{T}$, where $\hat{y}_{j,k}=q_j\left(y_{j,k}\right)$.

It can be seen that the consideration of miss detection and signal quantisation brings in extra nonlinearity and stochasticity in the addressed system, which makes the classical Kalman filter no longer applicable. Therefore, the RBPF, which can effectively deal with nonlinear and stochastic systems, is adopted to solve the STE and field reconstruction problems in the framework of Bayesian inference.

\subsection{RBPF approach} \label{subsec2-4}
Under the Bayesian framework, the estimation problem aims to derive the marginal posterior distribution $p(x_{k}| \hat{y}_{0:k})$ given the set of data $\hat{y}_{0:k}$ provided by the sensor network. To exploit the linear structure in the measurement model, analogue to \eqref{linearyk}, we introduce a latent variable $z_k$ defined as
\begin{align}\label{latentz}
    z_k = H_k x_{k},
\end{align}
and then we have
\begin{align} 
    p(x_{k}| \hat{y}_{0:k}) &= \int p(x_{k}|z_{0:k}) p(z_{0:k}|\hat{y}_{0:k}) \, \mathrm{d}z_{0:k}  
\end{align}
In Rao-Blackwellised particle filtering \citep{gelfand1990sampling,andrieu2002particle}, only the density $p(z_{0:k}|\hat{y}_{0:k})$ needs to be approximated using a Monte Carlo method, such that 
\begin{equation} 
\label{postMC}
    p\left(z_{0:k}|\hat{y}_{0:k}\right) \approx \sum_{m=1}^{M} \omega_{k}^{(m)} \delta_{z_{0:k}^{(m)}}(z_{0:k})
\end{equation}
where $\{z_{0:k}^{(m)}\}_{i=1}^{M}$ is the set of particles drawn from an importance sampling process related to $p(z_{0:k}|\hat{y}_{0:k})$, and $\omega_{k}^{(m)}>0$ are the corresponding weightings, with $\sum_{m=1}^{M} \omega_{k}^{(m)} = 1$. Then, the marginal posterior distribution can be approximated as 
\begin{equation}
    p(x_{k}| \hat{y}_{0:k}) \approx \sum_{m=1}^{M} \omega_{k}^{(m)} p(x_{k}| z_{0:k}^{(m)})
\end{equation}

As indicated above, RBPF aims to marginalise out part of the system state which is linear and Gaussian, such that classical Kalman filter can be applied to conduct Bayesian inference conditioned on the latent variable. Note that the miss detection in \eqref{missyk} is multiplicative and stochastic, and the quantisation in \eqref{quanti1} is nonlinear. As a result, neither $y_{j,k}$ nor $\hat{y}_{j,k}$ can be used as the desired latent variable in our setting. To solve the problem, $z_k$ in \eqref{latentz} is adopted as the latent variable, which is a linear and noise-free measurement with respect to the system state $x_k$. Therefore, $p \left( x_{k}|z_{k}^{(m)} \right)$ is Gaussian and can be obtained by a Kalman filter, if a Gaussian distribution with mean $ x_{k|k-1}^{(m)}$ and covariance matrix $P_{k|k-1}$ is associated with each particle $z_{k-1}^{(m)}$, at the previous sampling instant.

To realise the RBPF solution, we first provide the Kalman filter formulation and then introduce the sequential importance sampling process to construct the Monte Carlo approximation of the target density $p \left(z_{0:k}|\hat{y}_{0:k}\right)$ in \eqref{postMC}. 

\subsubsection{Kalman filter formulation}

The initial condition of each Kalman filter is set as $\hat{x}_{0|0}={\mathbb E}\left\{x_0\right\}$ and $P_{0|0}>0$. For time instant $k=0, 1,2,\ldots$, based on the system in Eq.~\eqref{xk+1}, the state prediction stage can be summarised as follows. 
\begin{align} \label{xpredict}
\hat{x}_{k+1|k} &= \bar{A}(p_{s,k})\hat{x}_{k|k}, \\
P_{k+1|k} &= \bar{A}(p_{s,k})P_{k|k}\bar{A}^\text{T}(p_{s,k})+\bar{W}_{k},
\end{align}
where $\hat{x}_{k+1|k}$ is the mean of the one-step prediction for both $\hat{c}_{k+1}$ and $u_{k+1}$, and $P_{k+1|k}$ is the prediction error covariance, defined as $$P_{k+1|k}\triangleq{\mathbb E}\left\{\left(x_{k+1}-\hat{x}_{k+1|k}\right)\left(x_{k+1}-\hat{x}_{k+1|k}\right)^\text{T}\right\}.$$

In the update stage, first set $z_{j,k}$ as the $j$th component of $z_k$, and it is clear that 
\begin{equation}
z_{j,k} = \bar{H}_{j} x_{k},
\end{equation} 
and the predicted value of $z_{j,k+1|k}$ is given as 
\begin{equation}
z_{j,k+1|k} = \bar{H}_{j} \hat{x}_{k+1|k}.
\end{equation}
Defining the innovation covariance for $z_{j,k+1}$ as
$$S_{j,k+1}=\bar{H}_{j} P_{k+1|k} \bar{H}_{j}^\text{T},$$
we obtain 
\begin{align}
    \label{yprob}
    p\left( z_{j,k+1}|z_{0:k}\right) = &  \mathcal{N}\left(z_{j,k+1|k}, S_{j,k+1}\right)
\end{align}

The update stage of the Kalman filter can now be constructed as follows
\begin{align} \label{update1}
K_{k+1} =& P_{k+1|k}H^\text{T}S_{k+1}^{-1},\\
\hat{x}_{k+1|k+1}=&\hat{x}_{k+1|k}+K_{k+1}\left(z_{k+1}-H\hat{x}_{k+1|k}\right), \label{estimate} \\
P_{k+1|k+1} =& \left(I-K_{k+1}H\right)P_{k+1|k},
\end{align}
where $$S_{k+1}=H P_{k+1|k} H^\text{T},$$ $K_{k+1}$ is the Kalman filter gain, and $\hat{x}_{k+1|k+1}$ and $P_{k+1|k+1}$ are the mean and covariance of the posterior Gaussian density, respectively. Note that $z_{k+1}$ in \eqref{estimate} is a new latent variable particle that is to be drawn from a proposal density as shown in the next subsection. 

\subsubsection{Sequential Monte Carlo approximation}
Following \eqref{postMC}, we aim to obtain the sequential Monte Carlo approximation of the density $p\left(z_{0:k}|\hat{y}_{0:k}\right)$ with a group of weighted particles. The importance sampling approach requires that this target density can be pointwise evaluated up to a normalising factor. We have 
\begin{equation} \label{target}
    p\left(z_{0:k}|\hat{y}_{0:k}\right) \propto \prod_{t=1}^{k} p\left( \hat{y_{t}} | z_{t} \right) p\left( z_{t} | z_{t-1}\right)
\end{equation}
where $p\left( z_{0} | z_{-1}\right) \triangleq p\left( z_{0} \right)$ is the prior distribution. Note that $p\left( \hat{y_{t}} | z_{t} \right)$ is known up to a normalisation constant by combining \eqref{linearyk} and \eqref{quanti1}, whereas $p\left( z_{t} | z_{t-1}\right)$ can be evaluated based on \eqref{yprob}. Therefore, we are able to design the sequential importance sampling and resampling process as follows. 

At time instant $k-1$, assume that we have a set of particles $\{ z_{0:k-1}^{(m)}\}_{m=1}^{M}$ that assembles $p\left(z_{0:k-1}|\hat{y}_{0:k-1}\right)$. At time $k$, for each particle $z_{0:k-1}^{(m)}$, we draw a new sample $z_{k}^{(m)}$ from a proposal distribution $q_{k}\left(z_{k}|z_{0:k-1}^{(m)},\hat{y}_{0:k}\right)$ and add it to the existing particles to form $z_{0:k}^{(m)}$. The unnormalised weight for this particle can be calculated by incorporating \eqref{target} and assuming the measurements are independent, such that
\begin{align}\label{weight}
    \bar{\omega}_{k}(z_{0:k}^{(m)}) 
    &\propto \frac{p\left(z_{0:k}^{(m)}|\hat{y}_{0:k}\right)}{q_{k}\left(z_{k}|z_{0:k-1}^{(m)},\hat{y}_{0:k}\right) p\left(z_{0:k-1}^{(m)}|\hat{y}_{0:k-1}\right)} \nonumber \\
    &\propto \frac{p\left(\hat{y}_{k}|z_{k}^{(m)}\right) p\left(z_{k}^{(m)} | z_{0:k-1}^{(m)} \right) }{q_{k}\left(z_{k}^{(m)}|z_{0:k-1}^{(m)},\hat{y}_{0:k}\right)}\nonumber \\
    &= \prod_{j=1}^N \frac{p\left(\hat{y}_{j,k}|z_{j,k}^{(m)}\right) p\left(z_{j,k}^{(m)} | z_{0:k-1}^{(m)} \right) }{q_{k}\left(z_{j,k}^{(m)}|z_{0:k-1}^{(m)},\hat{y}_{0:k}\right)}
\end{align}
and the normalised weight is $\omega_{k}^{(m)} = \bar{\omega}_{k}^{(m)}/\sum_{m=1}^{M} \bar{\omega}_{k}^{(m)}$.

In our work, for simplicity, we choose $$z_{j,k}^{(m)}\sim U\left[\hat{y}_{j,k}-\frac{\eta_j}{\zeta_j},\hat{y}_{j,k}+\frac{\eta_j}{\zeta_j}\right].$$ In this case, it can be seen that
\begin{align}\label{importance}
    q_{k}\left(z_{j,k}^{(m)}|z_{0:k-1}^{(m)},\hat{y}_{0:k}\right)=\frac{ \zeta_j }{2\eta_j}
\end{align}

Note that $p(z_{j,k}^{(m)} | z_{1:k-1}^{(m)} )$ has been provided in \eqref{yprob}, so it only remains to calculate $p(\hat{y}_{j,k}|z_{j,k}^{(m)})$ in \eqref{weight}. According to the occurrence of miss detection, $p(\hat{y}_{j,k}|z_{j,k}^{(m)})$  can be determined as
\begin{align}\label{likelihood1}
p(\hat{y}_{j,k}|z_{j,k}^{(m)})=\sum\limits_{s=0}^{1}{\rm Prob}\left(\alpha_{j,k}=s\right)p(\hat{y}_{j,k}|z_{j,k}^{(m)},\alpha_{j,k}=s).
\end{align}

Based on the analysis of the distribution of a quantised signal in \cite{liu2022novel}, we have
\begin{align}\label{likelihood2}
&p\left(\hat{y}_{j,k}|z_{j,k}^{(m)},\alpha_{j,k}=1\right)\nonumber\\
=&\Phi\left(\hat{y}_{j,k}+\frac{\eta_j}{\zeta_j}\bigg|z_{j,k}^{(m)},V_{j,k}\right)
-\Phi\left(\hat{y}_{j,k}-\frac{\eta_j}{\zeta_j}\bigg|z_{j,k}^{(m)},V_{j,k}\right),
\end{align}
and
\begin{align}\label{likelihood3}
&p\left(\hat{y}_{j,k}|z_{j,k}^{(m)},\alpha_{j,k}=0\right)\nonumber\\
=&\Phi\left(\hat{y}_{j,k}+\frac{\eta_j}{\zeta_j}\bigg|0,V_{j,k}\right)-\Phi\left(\hat{y}_{j,k}-\frac{\eta_j}{\zeta_j}\bigg|0,V_{j,k}\right),
\end{align}
where $\Phi(a|\mu,\sigma^2)$ is the cumulative distribution function of $a$ with respect to the Gaussian distribution $\mathcal{N}(\mu,\sigma^2)$.

With Eqs.~\eqref{likelihood1}, \eqref{likelihood2} and \eqref{likelihood3}, $p(\hat{y}_{j,k}|z_{j,k}^{(m)})$ can be calculated, and the particle weight can be updated with \eqref{weight}.

Resample the particles with the classical multinomial resampling method, and then the state estimate can be obtained as
\begin{align}\label{esti}
\bar{x}_k=\sum\limits_{m=1}^M \omega_{k}^{(m)} x_{k|k}^{(m)},
\end{align}
where $x_{k|k}^{(m)}$ is determined with \eqref{update1}.

\begin{figure*}[htbp]
\begin{center}
  \includegraphics[width=0.9\linewidth]{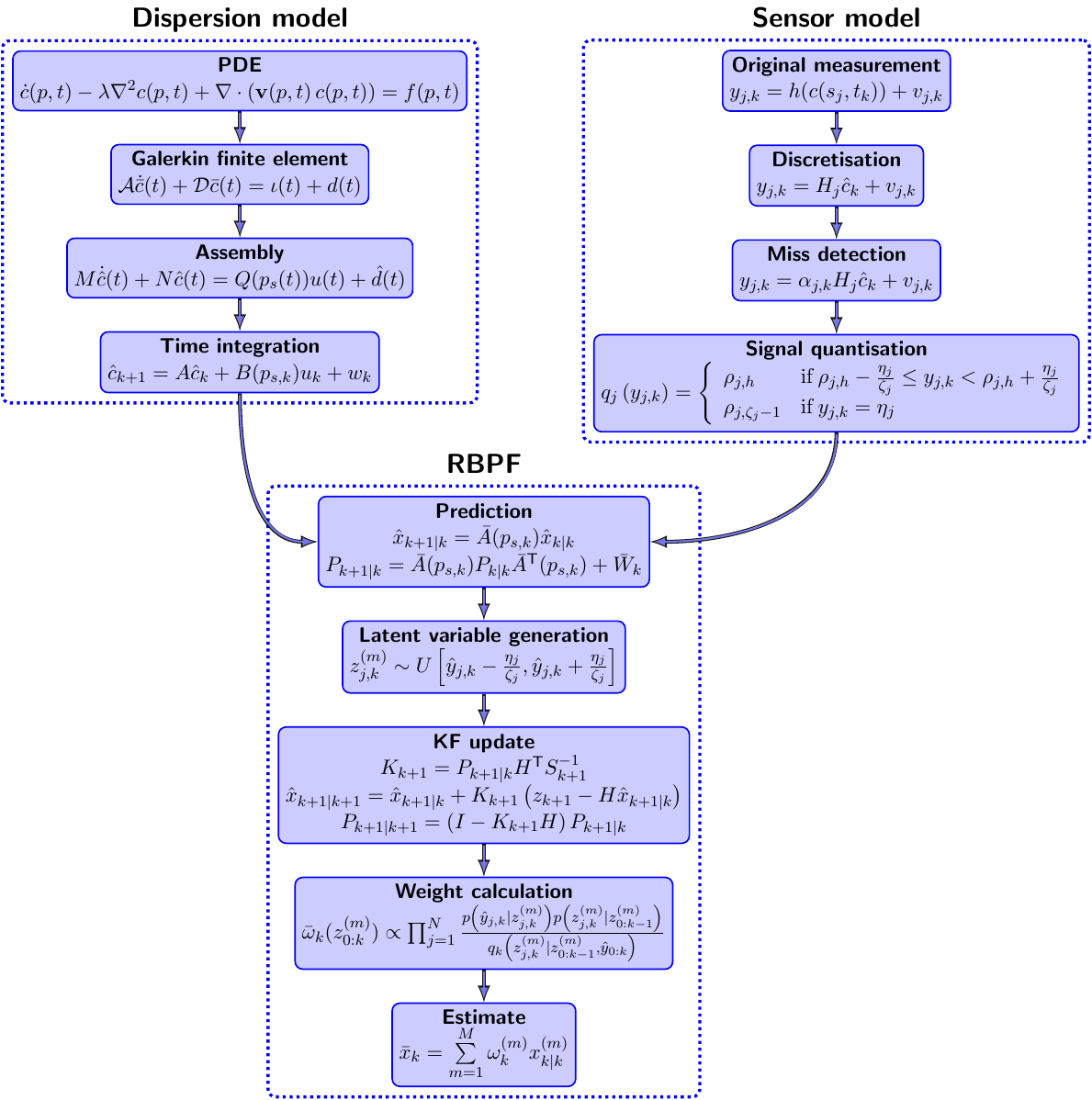}
\caption{Flowchart of overall algorithm}\label{flowchart}
\end{center}
\end{figure*}

The overall algorithm is summarised in Algorithm \ref{ConDPF}. To better visualise the overall algorithm, the flowchart is given in Figure \ref{flowchart}.
\begin{algorithm}[t]
\caption{The RBPF-based source strength and dispersion estimation at the $k$th time step} \label{ConDPF}
\LinesNumbered
1.~~Approximate and discretise Eq.~\eqref{PDE1} as Eq.~\eqref{xk+1} with Eqs.~\eqref{Galerkin8}, \eqref{stiffness_matrix2}--\eqref{force_vector3}, \eqref{para1} and \eqref{mass_matrix3}.\\
2.~~Collect $\hat{y}_{i,k}$ with Eqs.~\eqref{missyk} and \eqref{quanti1}. \\
3.~~Update $P_{k+1|k}$ and $P_{k+1|k+1}$ with Eqs.~\eqref{xpredict} and \eqref{update1}.\\
\For{$m=1,2,\cdots,M$} {4.~~Generate prediction $x_{k|k-1}^{(m)}$ with Eq.~\eqref{xpredict}. \\
5.~~Draw $z_{j,k}^{(m)}\sim U\left[\hat{y}_{j,k}-\eta_j/ \zeta_j,\hat{y}_{j,k}+\eta_j/ \zeta_j\right]$. \\
6.~~Determine $\bar{\omega}_{k}^{(m)}$ with Eqs.~\eqref{yprob}, \eqref{weight},  \eqref{importance}, and \eqref{likelihood1}.\\
7.~~Calculate $x_{k|k}^{(m)}$ with Eq.~\eqref{update1}. \\
}
8.~~Normalise the weights and resample with the multinomial resampling technique.\\
9.~~Obtain $\bar{x}_k$ with Eq.~\eqref{esti}.
\end{algorithm}

\section{Case study} \label{sec3}
In this section, an oil leakage example is designed based on the Baltic current data. The effectiveness of the proposed modelling and RBPF algorithms will be demonstrated and the comparative results will be presented with respect to a benchmark ensemble Kalman filer (EnKF). The simulations were performed on a 2.90 GHz desktop computer with an Intel(R) Core(TM) i7-7500 CPU, 16.0GB RAM, 64-bit operating system, and x64-based processor. The software was R2020b 64-bit MATLAB.
 
\subsection{Characteristics of the system setting} \label{subsec3-1}

In the following simulation experiments, we considered the diffusion of crude oil in ocean environments following an oil spill incident. In this case, the diffusion coefficient $\lambda$ in Eq.~\eqref{PDE1} is $1.12\times 10^{-8}\,\text{m}^2\,\text{s}^{-1}$ \citep{hamam1987diffusion}. The coordinate of source was set as $(6.129, 12.06)\times 10^{5}\,\text{m}$. The decay rate was zero, and the source emission strength was $1\,\text{g}\,\text{m}^{-3}\,\text{s}^{-1}$.

\begin{figure*}
\begin{center}
  \includegraphics[width=0.75\linewidth]{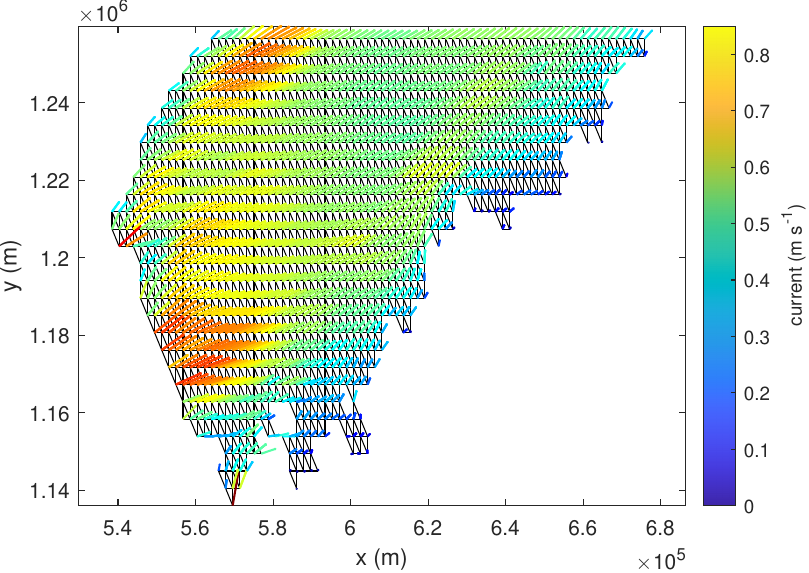}
\caption{Finite element mesh and current map at the first sampling instant}\label{current}
\end{center}
\end{figure*}

The ocean current plays a crucial role in the dispersion of hazardous materials. In this paper, we adopt real Baltic current data, which can be found on \href{https://www.tidetech.org/data/}{https://www.tidetech.org/data/}. The data were recorded from 2017-08-20 00:00 to 2017-08-22 12:00 once per hour, and covered the region from $9.04^{\circ}$E to $30.24^{\circ}$E and from $53.02^{\circ}$N to $65.83^{\circ}$N with a resolution of $0.040^{\circ}$. The coordinates of the sampling points are distances with respect to the origin point at $0^{\circ}$E, $0^{\circ}$N. The data set includes both latitudinal and longitudinal current speeds with units of $\text{m}\,\text{s}^{-1}$. Fig.~\ref{current} illustrates the current map at the first sampling time and the finite element mesh of the ROI, which can help us to have a visual understanding of the data set. In our investigation, the ROI is between the coordinates $(5.3081,6.8354)\times 10^{5}\,\text{m}$ and $(1.1361,1.256)\times 10^{6}\,\text{m}$
where the coastline effect on the dispersion process is included.

\begin{figure*}[htbp]
	\centering
	\subfigure[$k=6$]{
		\begin{minipage}[b]{0.4\textwidth}
			\includegraphics[width=\textwidth]
			{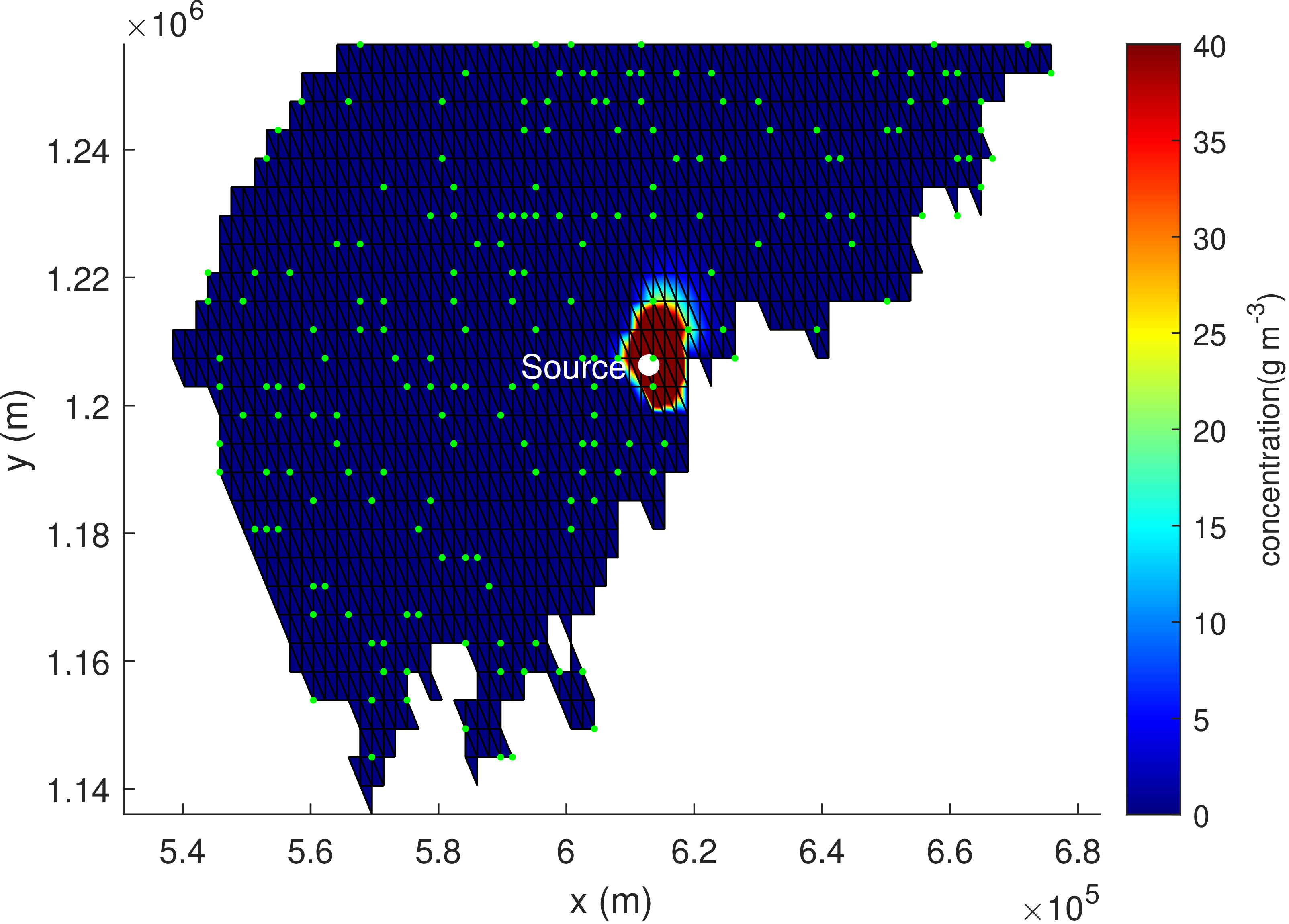}\label{field1}\end{minipage}
	}
    	\subfigure[$k=12$]{
    		\begin{minipage}[b]{0.4\textwidth}
   		 	\includegraphics[width=\textwidth]
   		 	{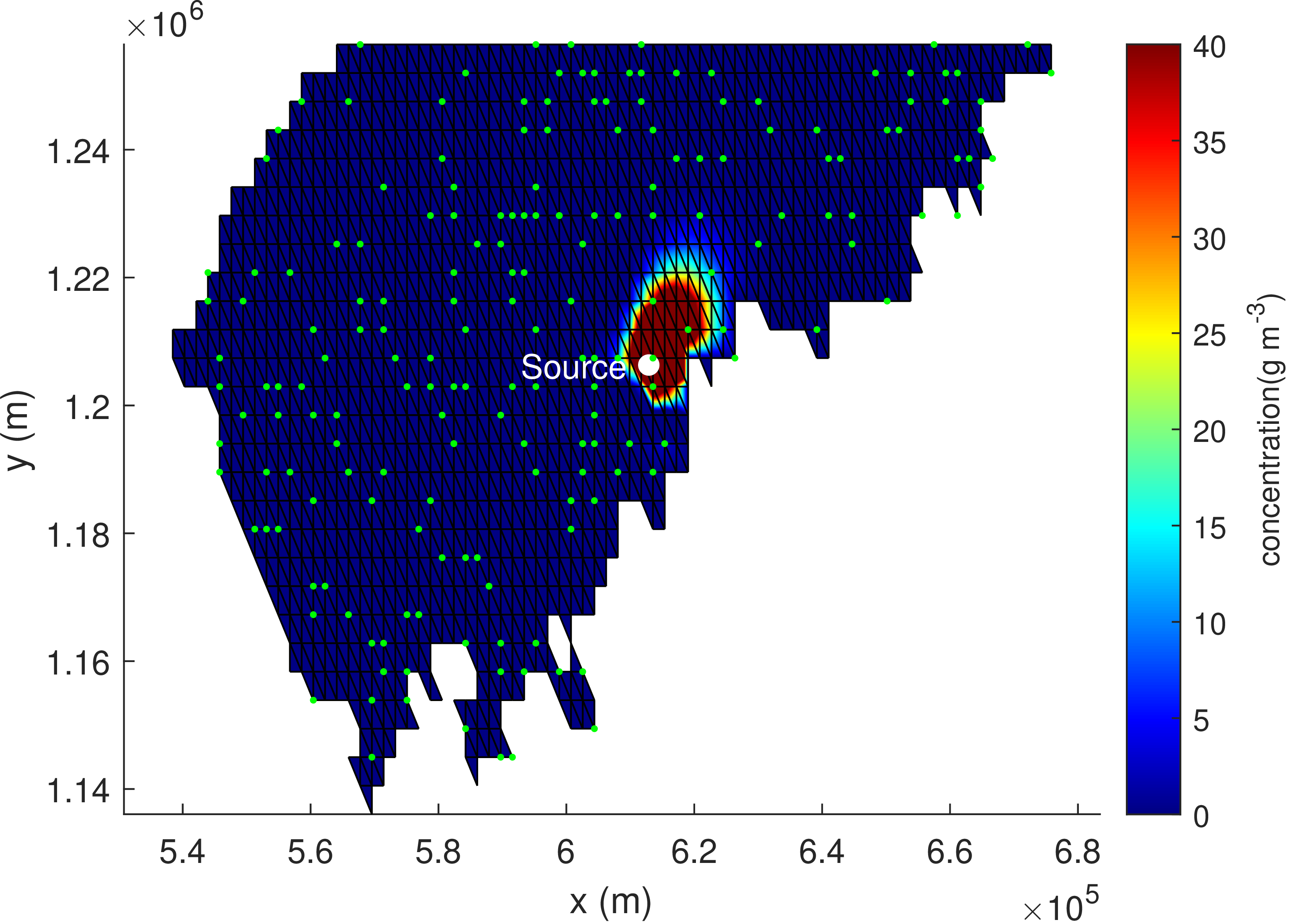}\label{field2}\end{minipage}
    	}
	\subfigure[$k=18$]{
		\begin{minipage}[b]{0.4\textwidth}
			\includegraphics[width=\textwidth]
			{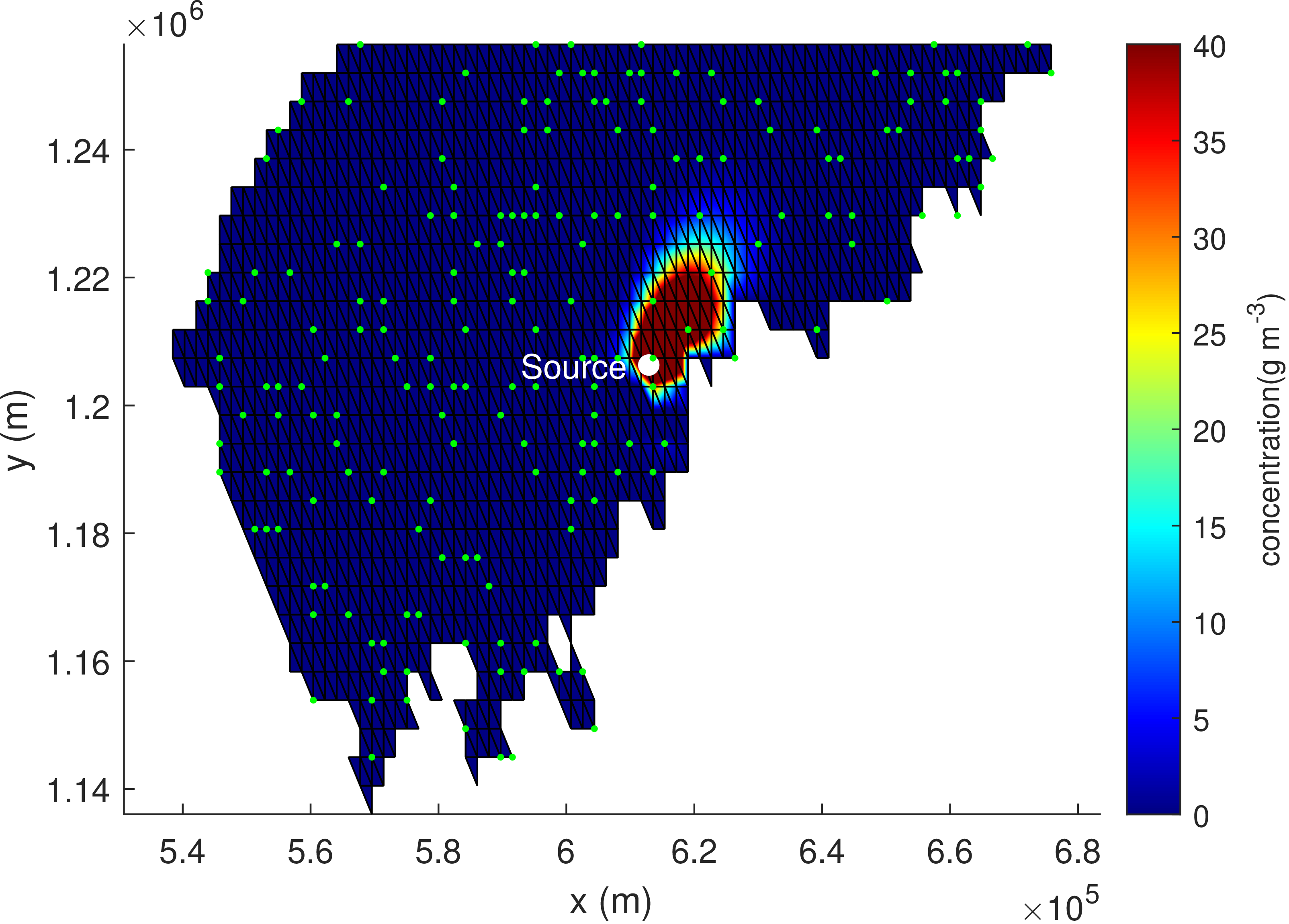}\label{field3}\end{minipage}
	}
    	\subfigure[$k=24$]{
    		\begin{minipage}[b]{0.4\textwidth}
		 	\includegraphics[width=\textwidth]
		 	{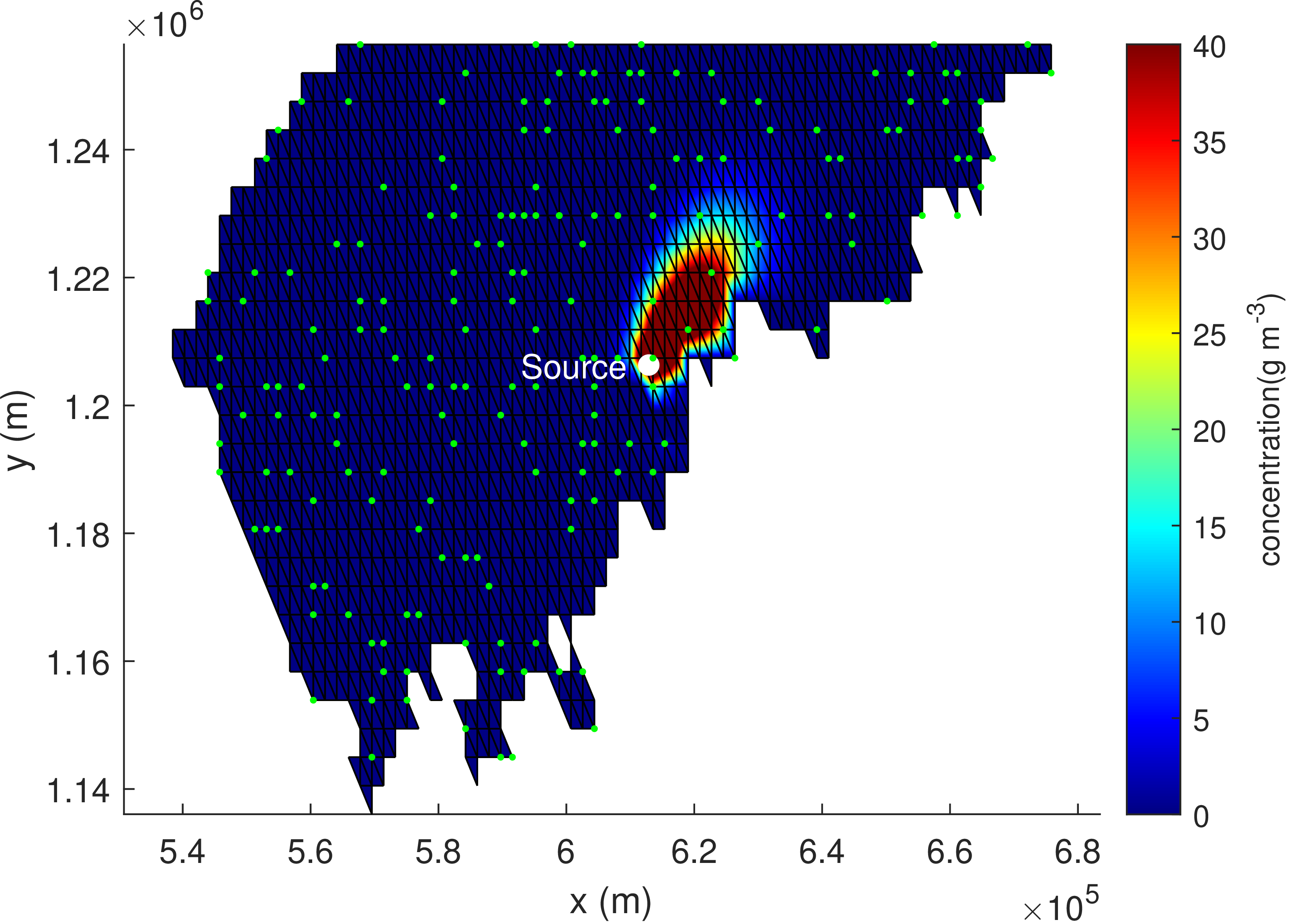}\label{field4}\end{minipage}
    	}
    		\\ 
	\subfigure[$k=30$]{
		\begin{minipage}[b]{0.4\textwidth}
			\includegraphics[width=\textwidth]
			{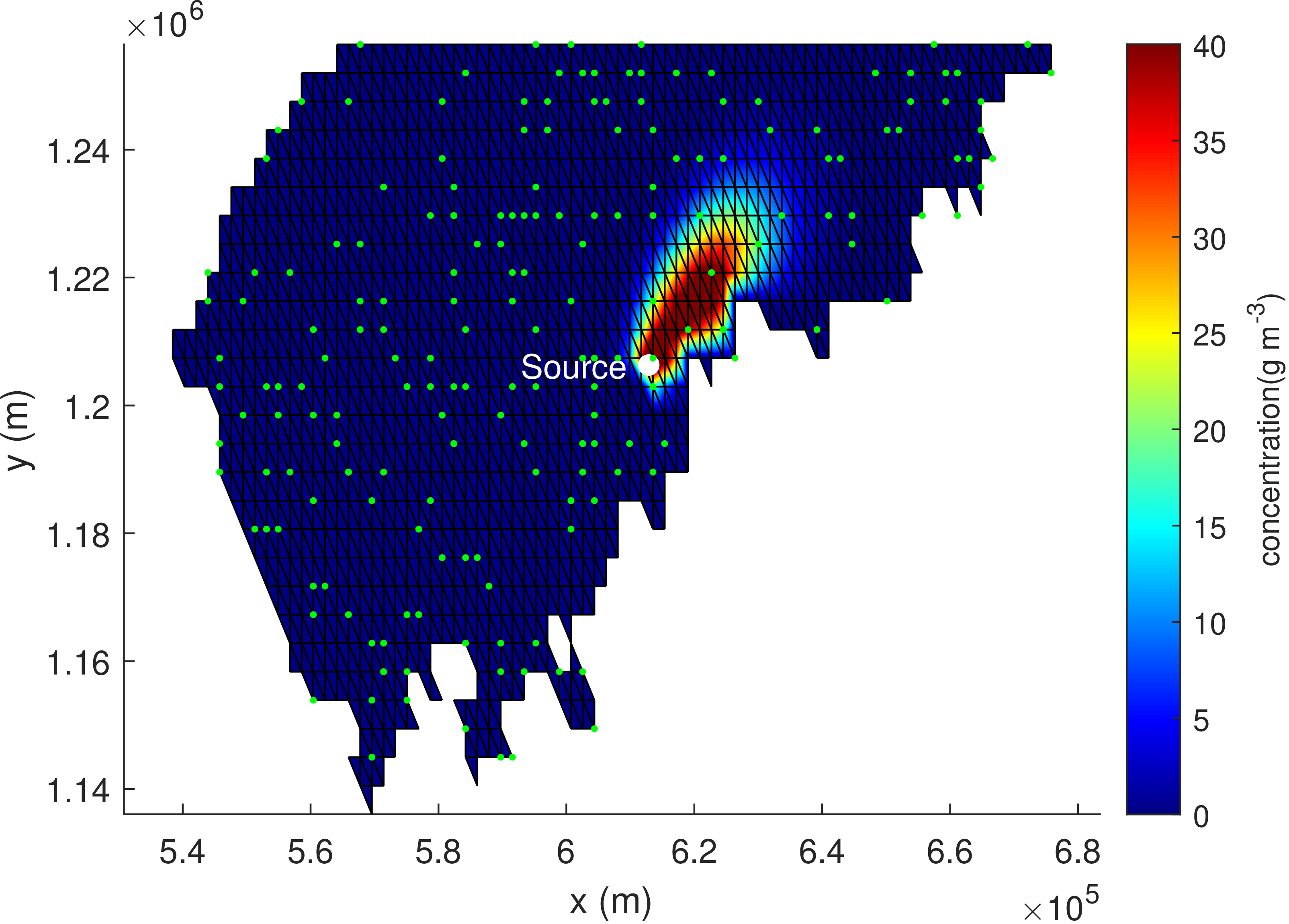}\label{field5}\end{minipage}
	}
    	\subfigure[$k=36$]{
    		\begin{minipage}[b]{0.4\textwidth}
   		 	\includegraphics[width=\textwidth]
   		 	{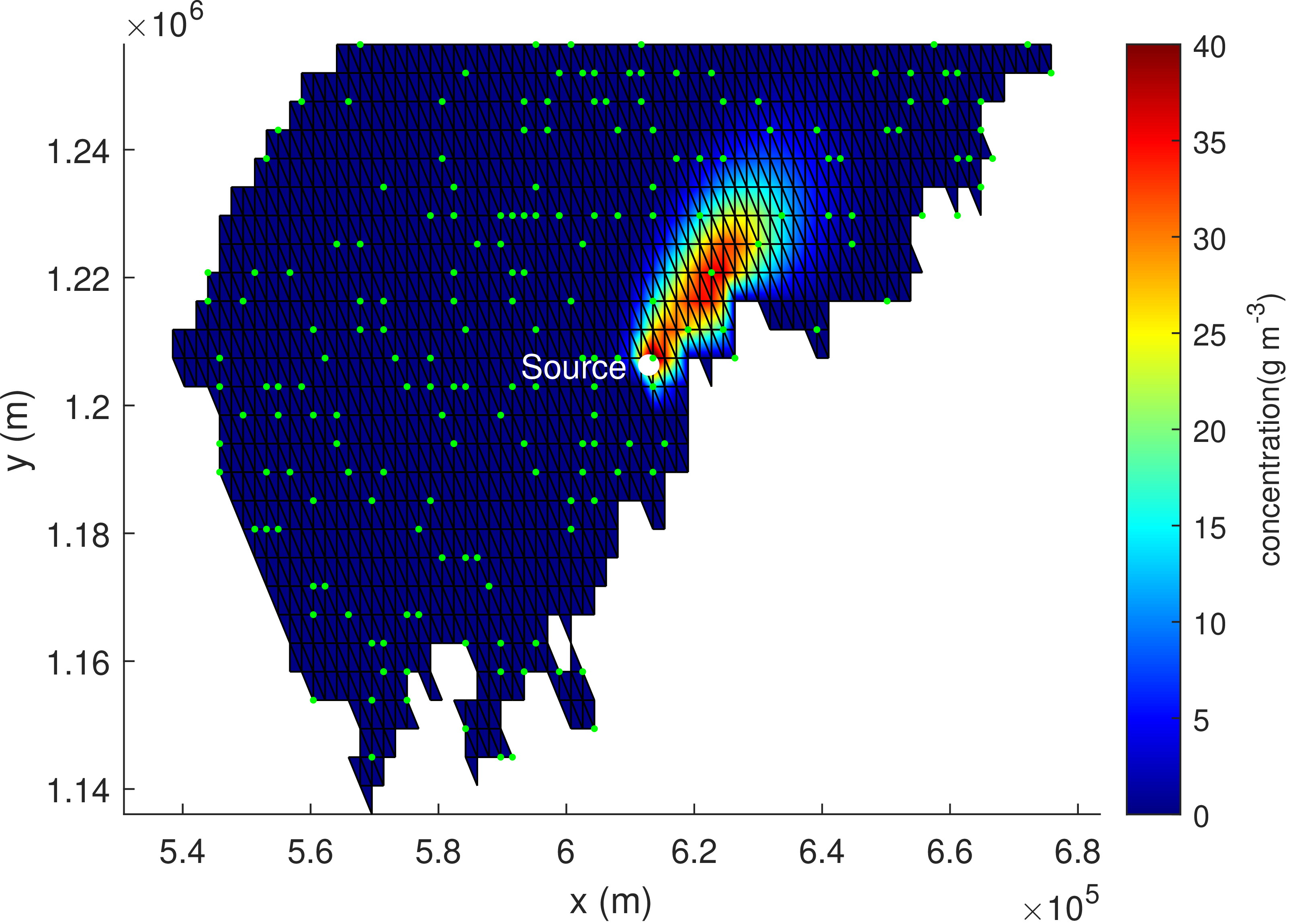}\label{field6}\end{minipage}
    	}
	\subfigure[$k=42$]{
		\begin{minipage}[b]{0.4\textwidth}
			\includegraphics[width=\textwidth]
			{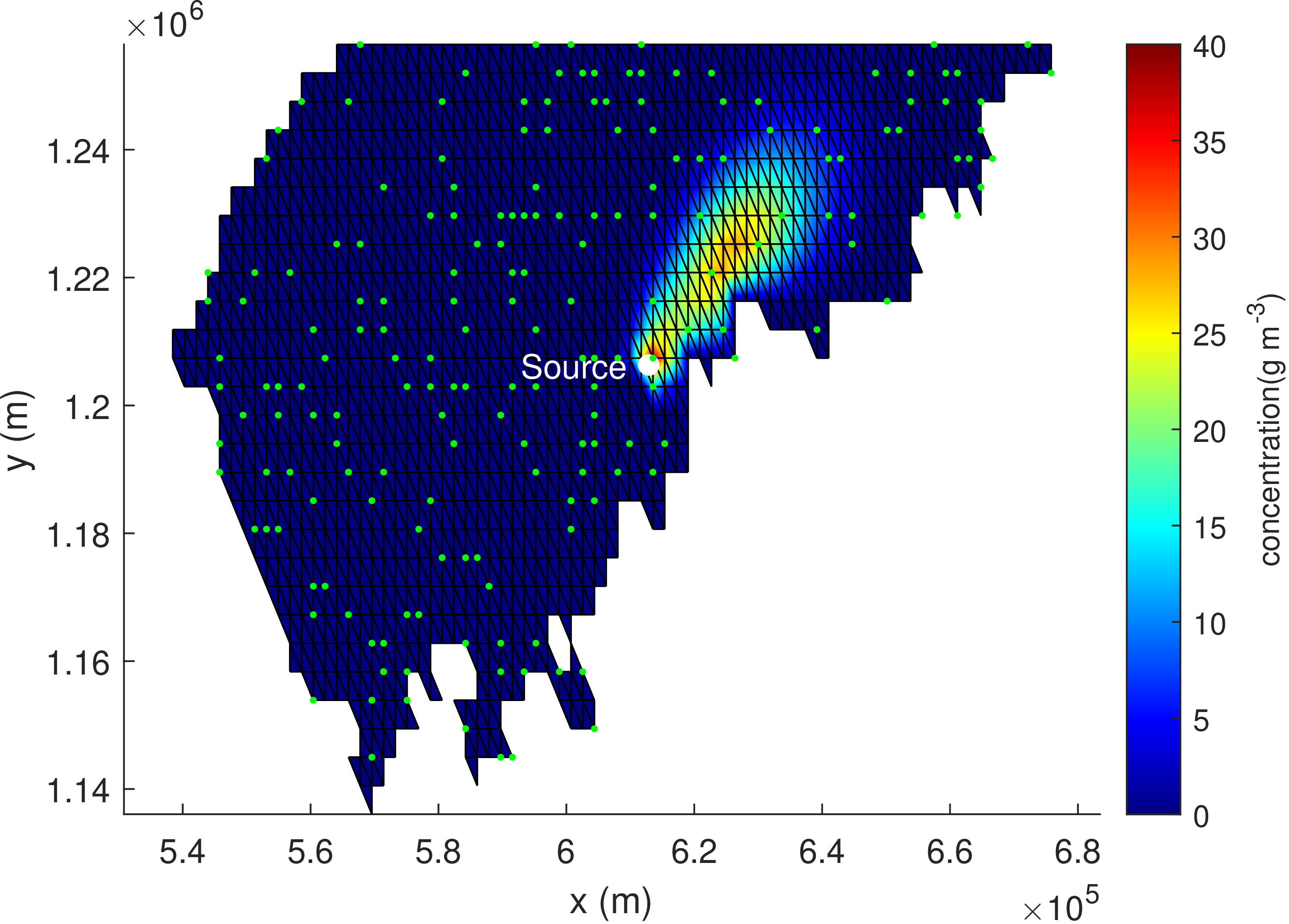}\label{field7}\end{minipage}
	}
    	\subfigure[$k=48$]{
    		\begin{minipage}[b]{0.4\textwidth}
		 	\includegraphics[width=\textwidth]
		 	{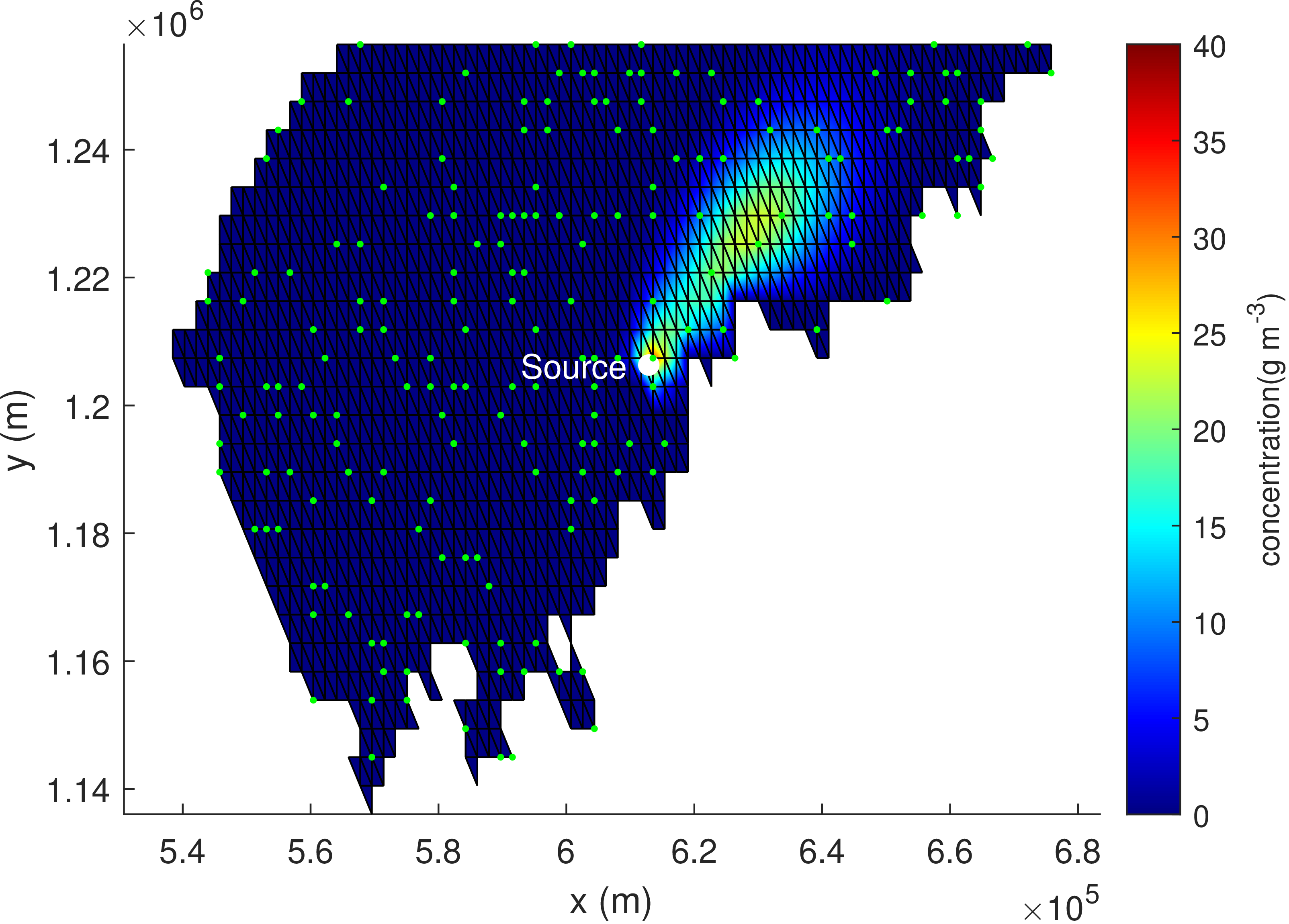}\label{field8}\end{minipage}
    	}
	\caption{Simulated ground-truth dispersion process at different time instants. The green dots represent the sensor locations}
	\label{field}
\end{figure*}

Consider 200 static sensors in the ROI which can collect and transmit the concentration measurements once per hour. The locations of the sensors were generated randomly by resorting to the uniform distribution. Both the dispersion and measurement noise $w_k$ and $v_{j,k}$ were zero-mean Gaussian distributed with covariance $5\times 10^{-3}I$. For the imperfect measurement phenomena, the miss detection rate $\bar{\alpha}_{j,k}=0.85$, the scaling parameter $\eta_j=660$, and the positive scalar $\zeta_j=1.1 \times 10^{4}$ for all the sensor $j$ and sampling step $k$. The initial guess for the source strength and dispersion process were all zero. The dispersion fields in the region at eight sampling instant ($k=6,12,\ldots,48$) were depicted in Fig.~\ref{field}, where the green dots represent the sensor locations.

\subsection{Estimation results of RBPF} \label{subsec3-2}

\begin{figure}
\begin{center}
  \includegraphics[width=0.6\linewidth]{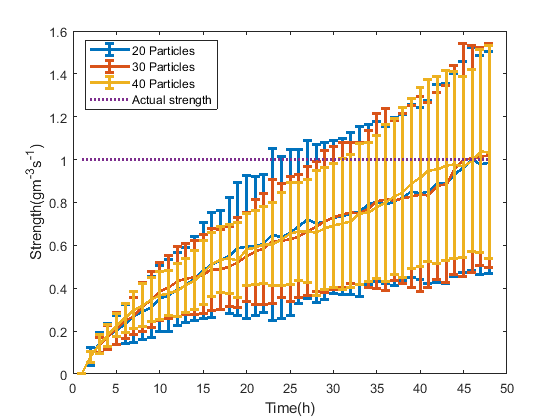}
\caption{Source strength estimation results with 20, 30 and 40 particles in RBPF. The error bar shows one standard deviation w.r.t. the Monte Carlo average.}\label{RBPF_source}
\end{center}
\end{figure}

The performance of the RBPF algorithm is first tested using Monte Carlo simulations. Fig.~\ref{RBPF_source} shows the average estimates of the emission strength obtained with different numbers of particles (i.e., $20$, $30$ and $40$) from 50 Monte Carlo trials. In light of this figure, the source strength can be well estimated with a smooth transition process in the presence of the noises and imperfect measurements. Although the convergence of the estimates is similar with different particle numbers in RBPF, it can be observed that the standard deviation with respect to the Monte Carlo average reduces, with more particles used in the filter, indicating that the consistence of the estimation can be improved with a larger particle numbers.   

\begin{figure*}[htbp]
	\centering
	\subfigure[$k=6$]{
		\begin{minipage}[b]{0.4\textwidth}
			\includegraphics[width=\textwidth]
			{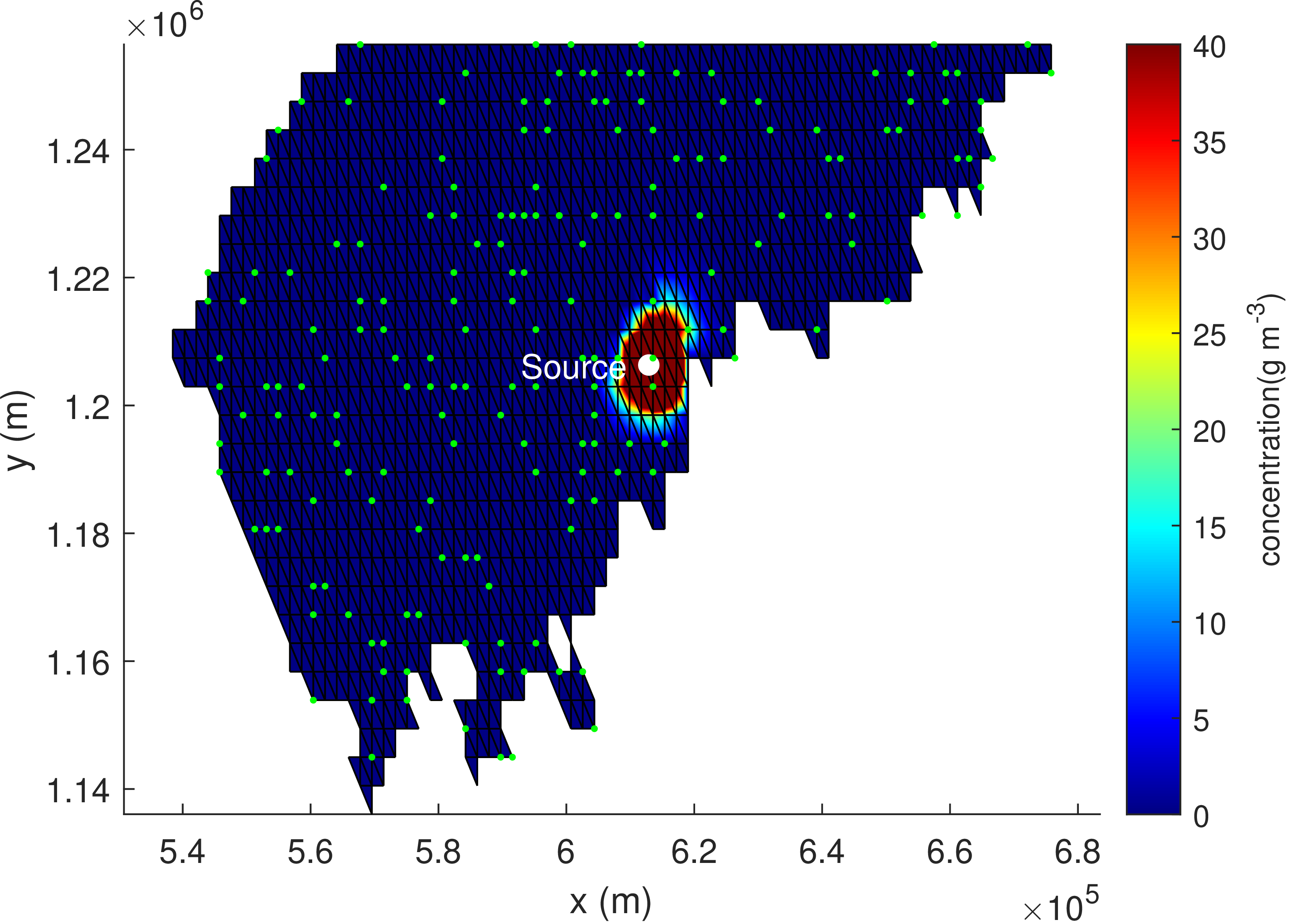}\label{RBPF_estimate1}\end{minipage}
	}
    	\subfigure[$k=12$]{
    		\begin{minipage}[b]{0.4\textwidth}
   		 	\includegraphics[width=\textwidth]
   		 	{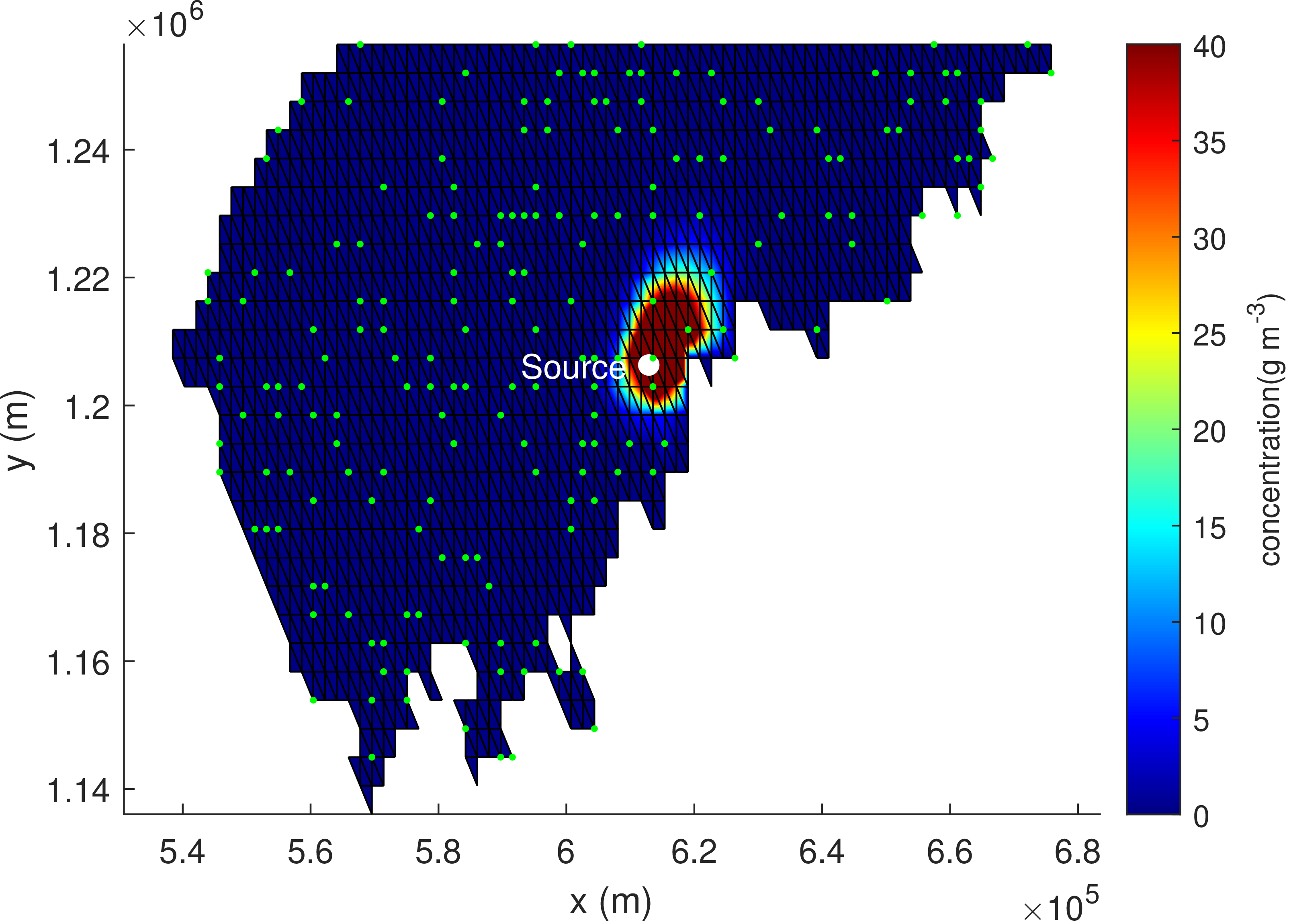}\label{RBPF_estimate2}\end{minipage}
    	}
	\subfigure[$k=18$]{
		\begin{minipage}[b]{0.4\textwidth}
			\includegraphics[width=\textwidth]
			{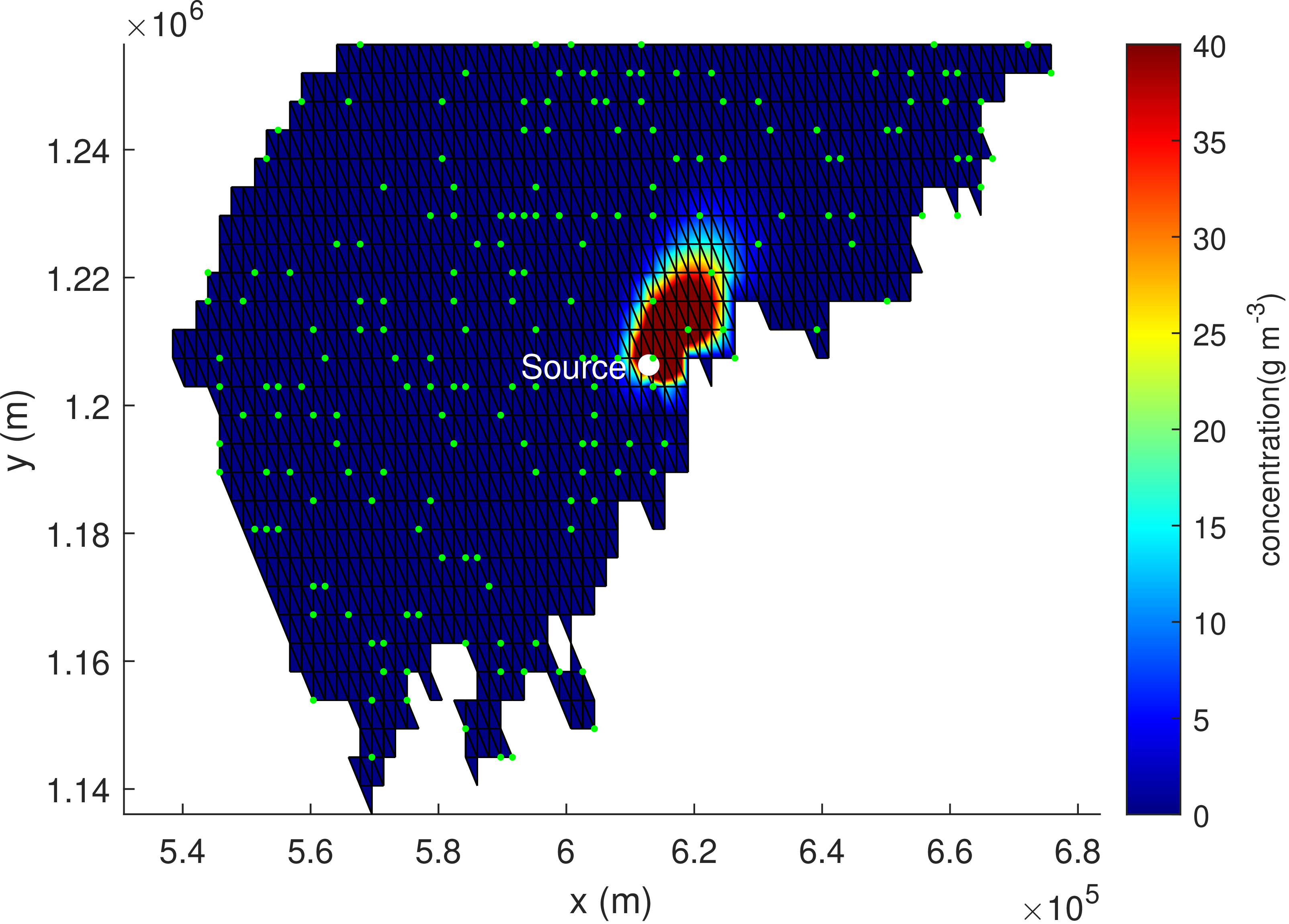}\label{RBPF_estimate3}\end{minipage}
	}
    	\subfigure[$k=24$]{
    		\begin{minipage}[b]{0.4\textwidth}
		 	\includegraphics[width=\textwidth]
		 	{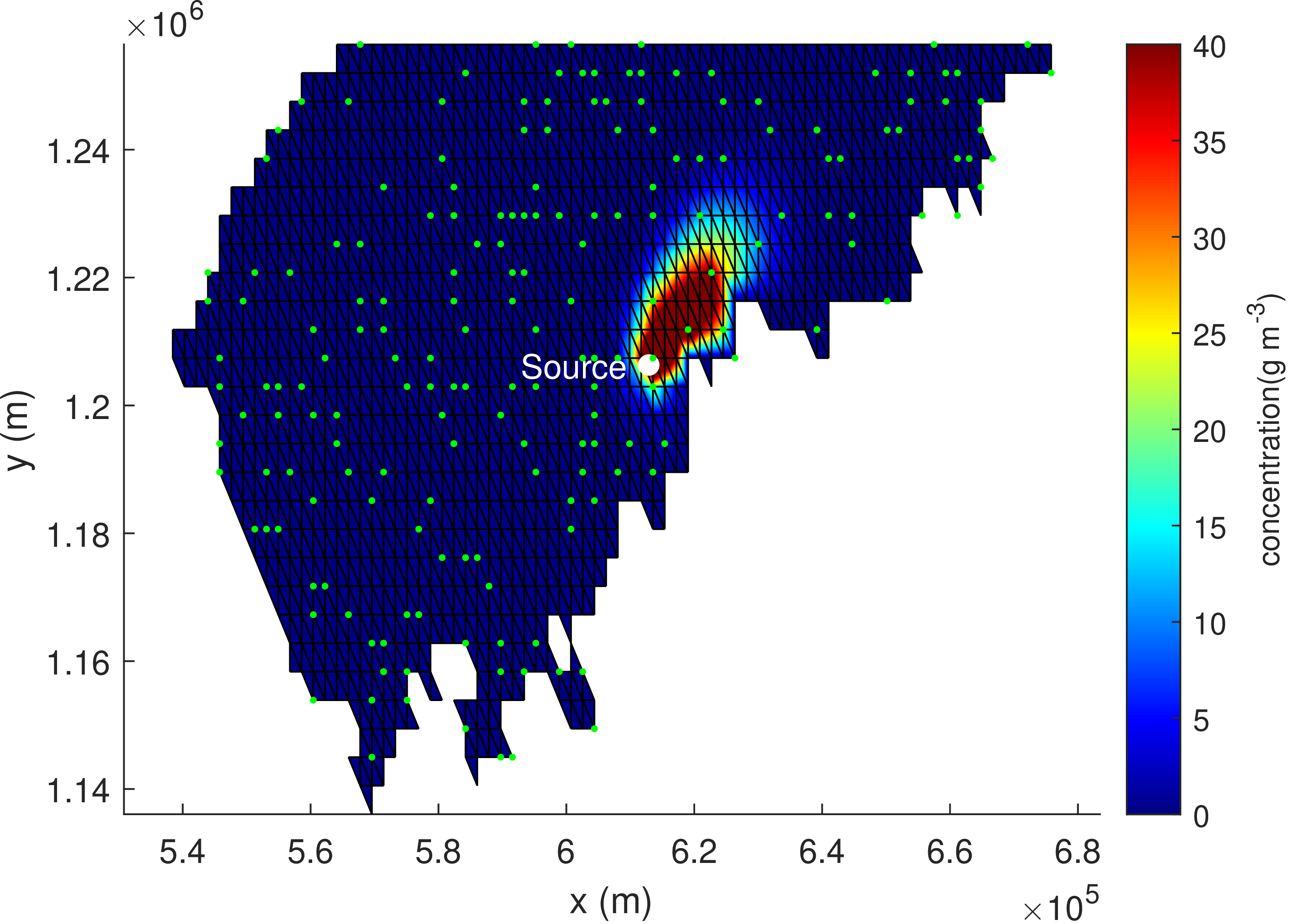}\label{RBPF_estimate4}\end{minipage}
    	}
    		\\ 
	\subfigure[$k=30$]{
		\begin{minipage}[b]{0.4\textwidth}
			\includegraphics[width=\textwidth]
			{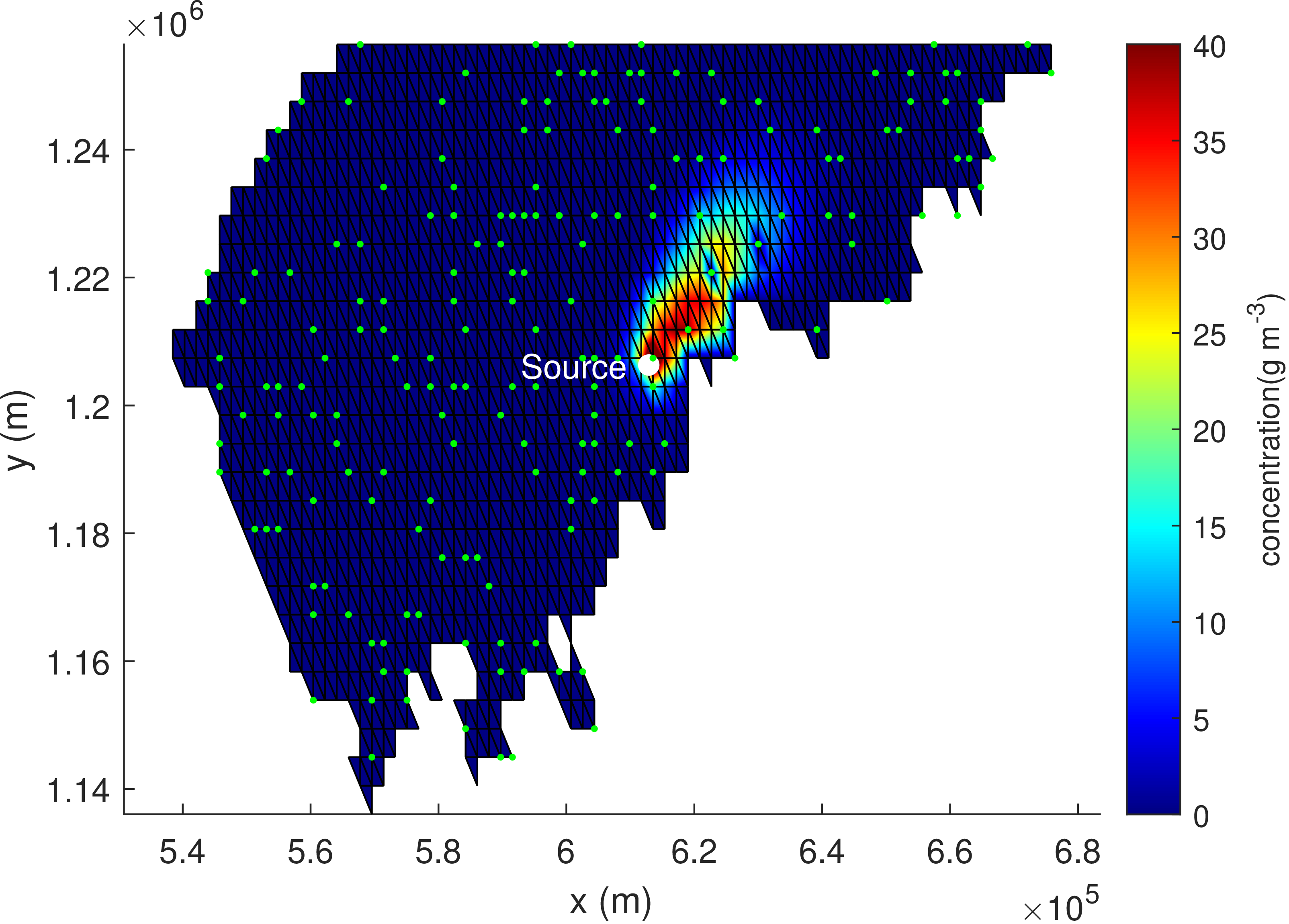}\label{RBPF_estimate5}\end{minipage}
	}
    	\subfigure[$k=36$]{
    		\begin{minipage}[b]{0.4\textwidth}
   		 	\includegraphics[width=\textwidth]
   		 	{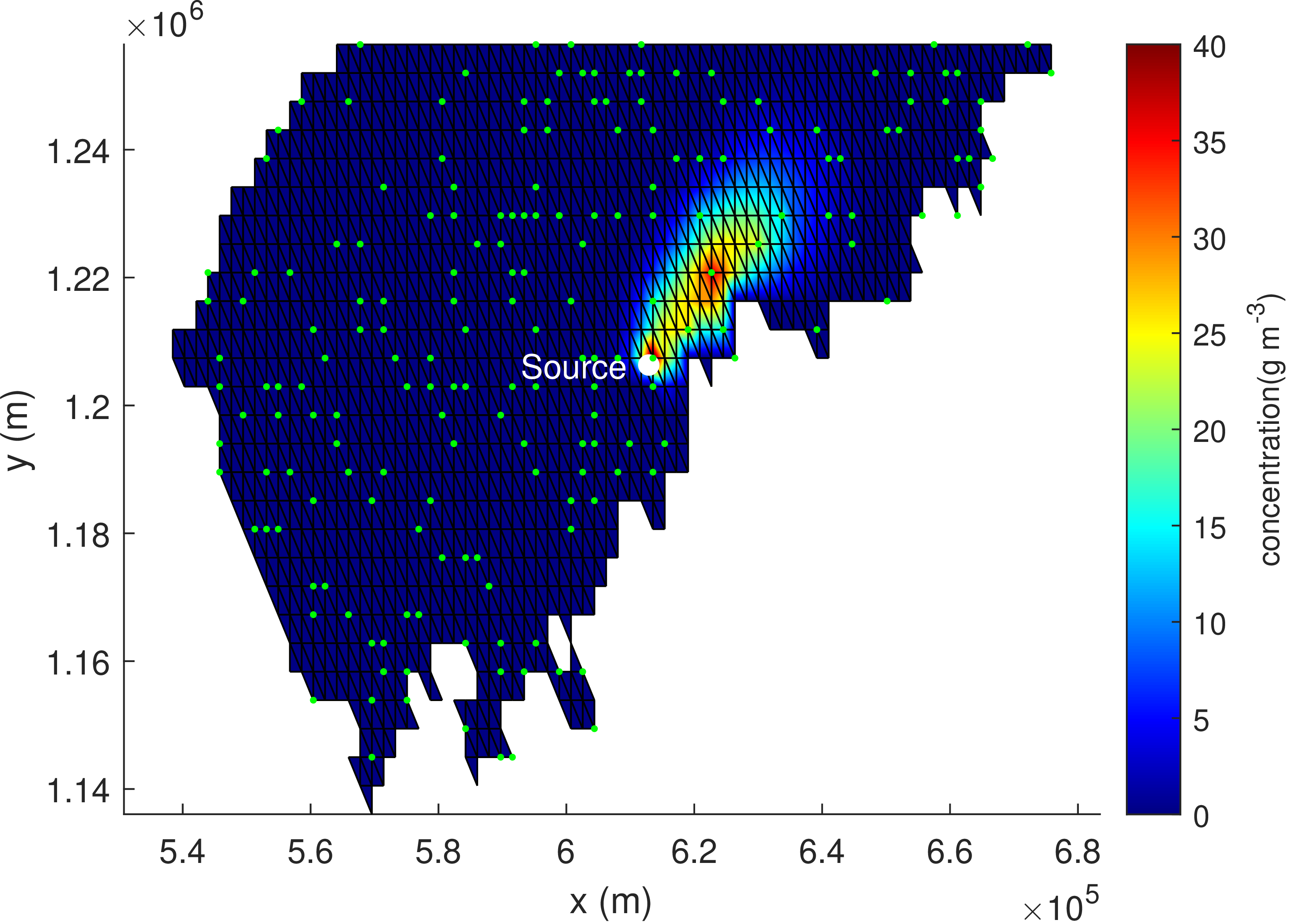}\label{RBPF_estimate6}\end{minipage}
    	}
	\subfigure[$k=42$]{
		\begin{minipage}[b]{0.4\textwidth}
			\includegraphics[width=\textwidth]
			{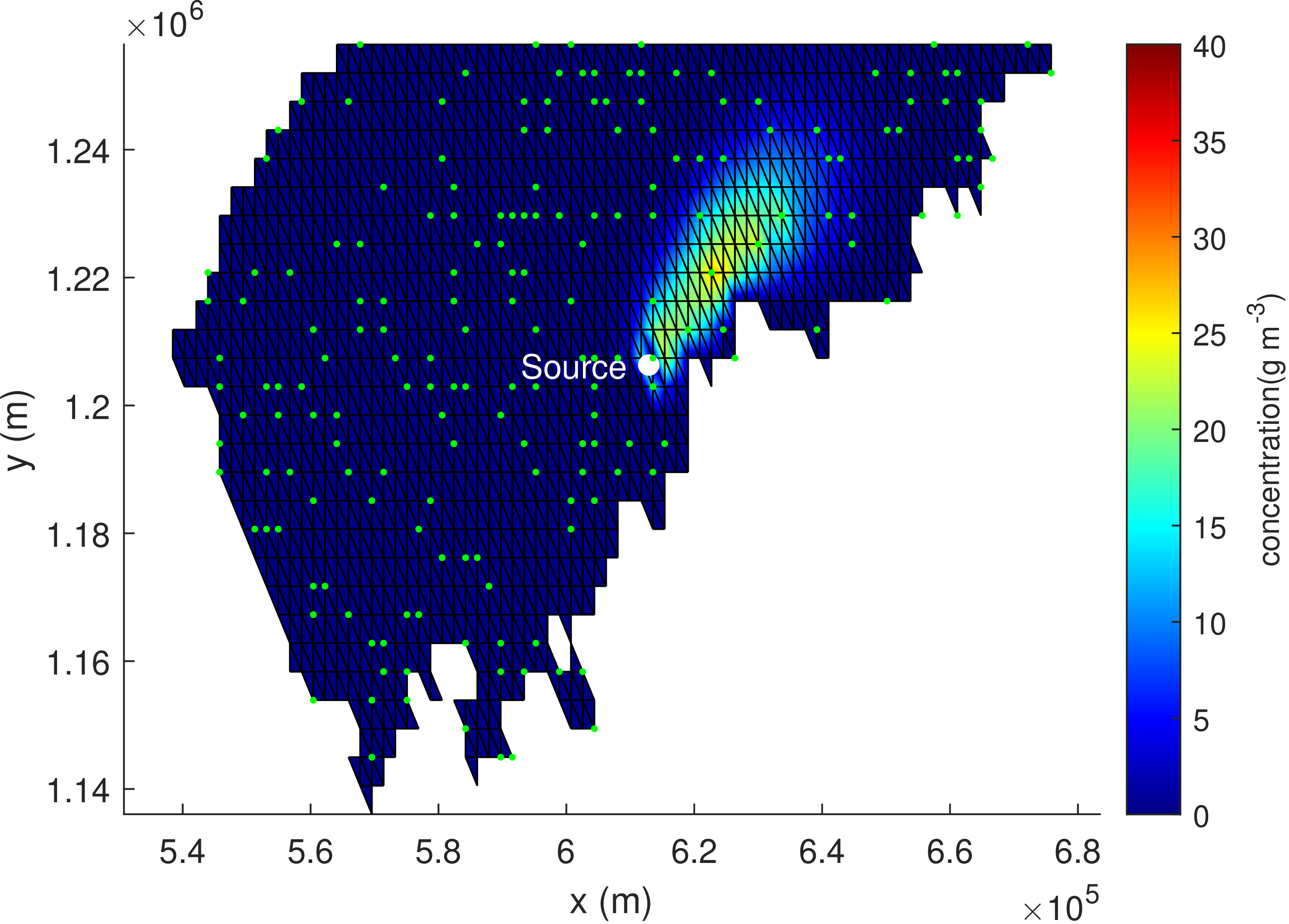}\label{RBPF_estimate7}\end{minipage}
	}
    	\subfigure[$k=48$]{
    		\begin{minipage}[b]{0.4\textwidth}
		 	\includegraphics[width=\textwidth]
		 	{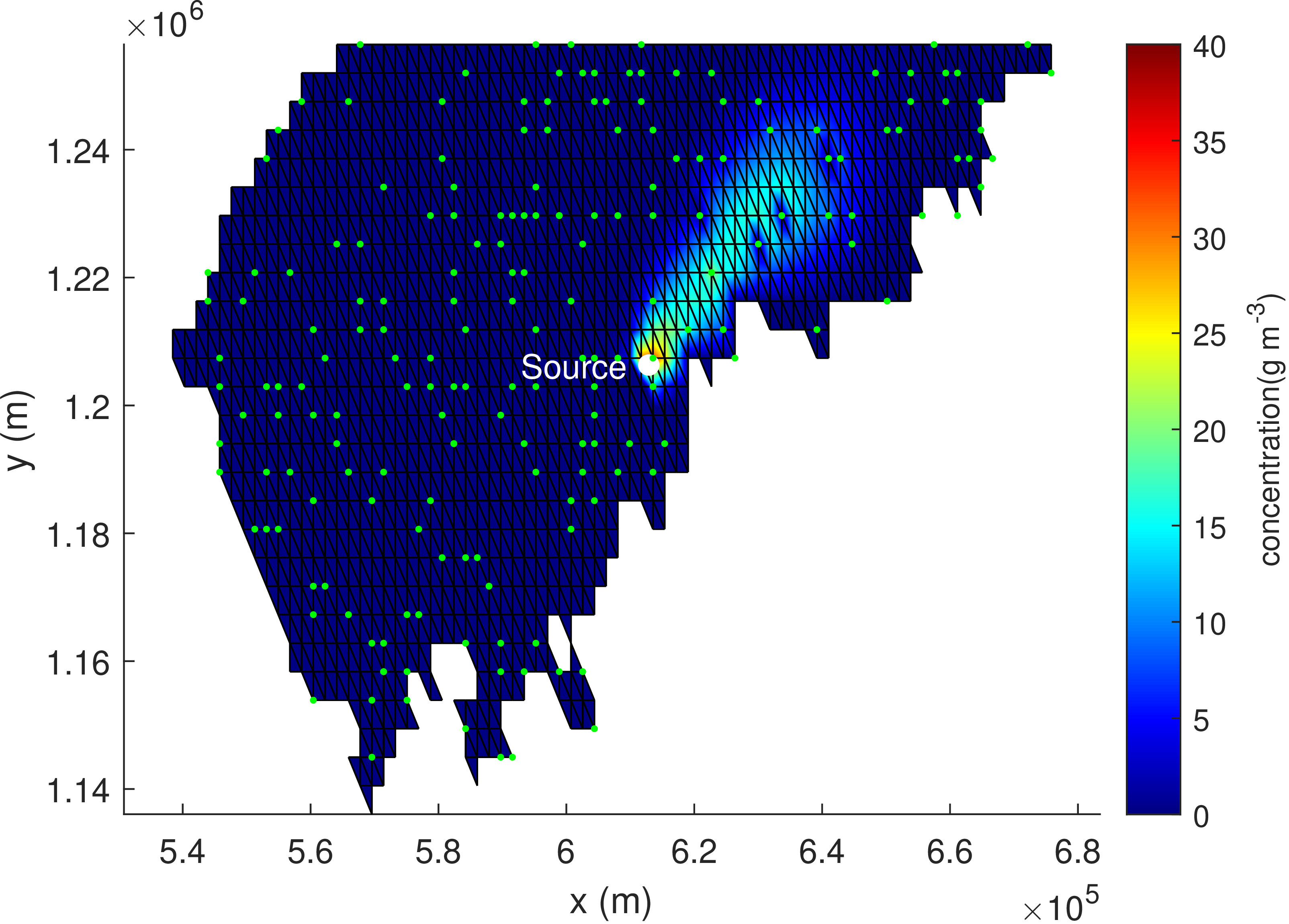}\label{RBPF_estimate8}\end{minipage}
    	}
	\caption{Dispersion process estimates with 30-particle RBPF at different sampling instants}
	\label{RBPF_estimate}
\end{figure*}

To give a qualitative illustration of the dispersion reconstruction capability, the results from one of the simulation runs is shown in Fig.~\ref{RBPF_estimate}, with the estimates of the dispersion process at different sampling instants. From Figs.~\ref{field} and \ref{RBPF_estimate}, it can be seen that the spatio-temporal dispersion process can be effectively reconstructed with the proposed RBPF. Even for such a high-dimensional problem, excellent estimation performance can be achieved with only 30 particles, which illustrates the effectiveness of RBPF in the reduction of particle number.

\subsection{Comparison study} \label{subsec3-3}

\begin{figure*}
	\centering
	\subfigure[$k=6$]{
		\begin{minipage}[b]{0.4\textwidth}
			\includegraphics[width=\textwidth]
			{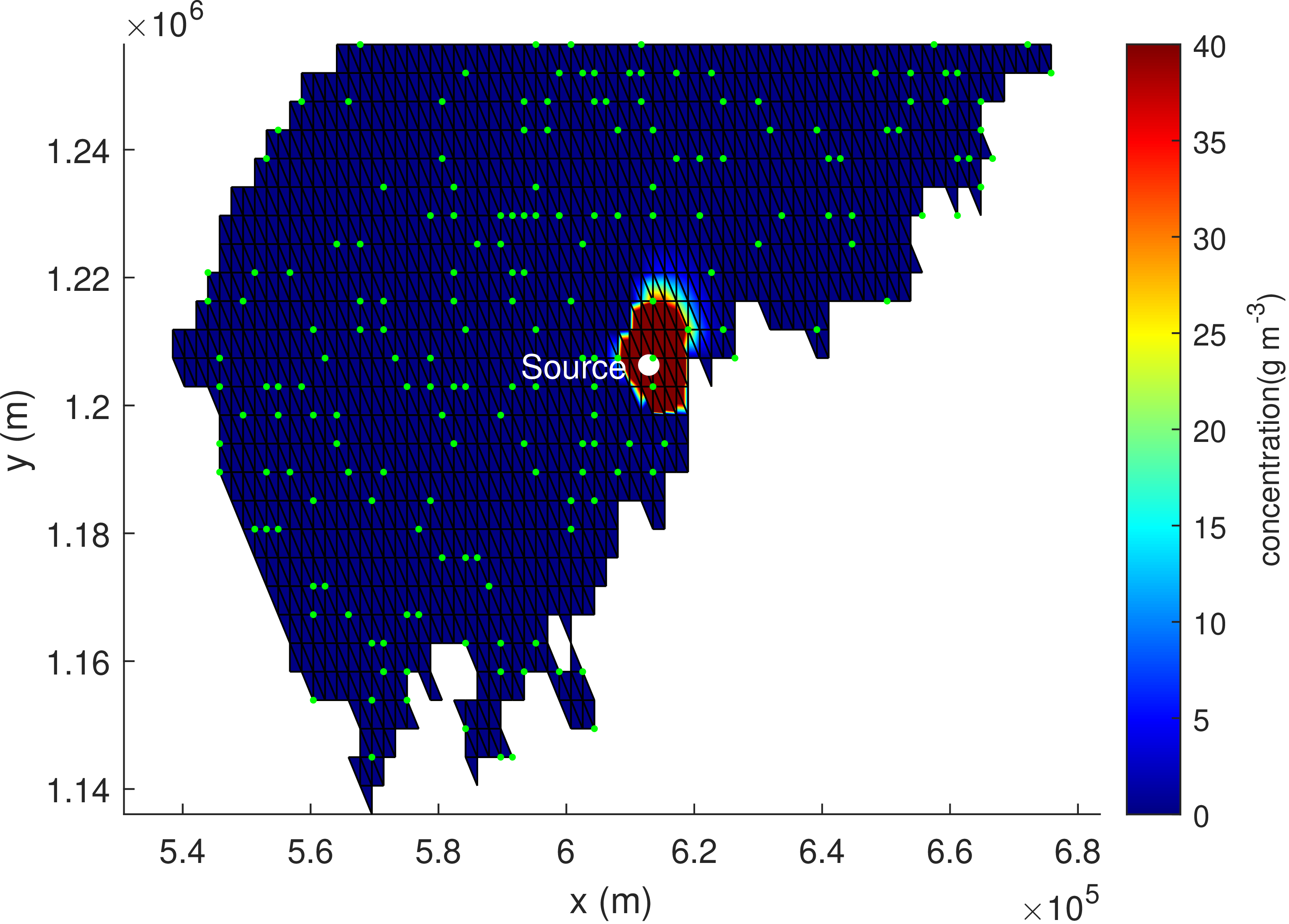}\label{EnKF_estimate1}\end{minipage}
	}
    	\subfigure[$k=12$]{
    		\begin{minipage}[b]{0.4\textwidth}
   		 	\includegraphics[width=\textwidth]
   		 	{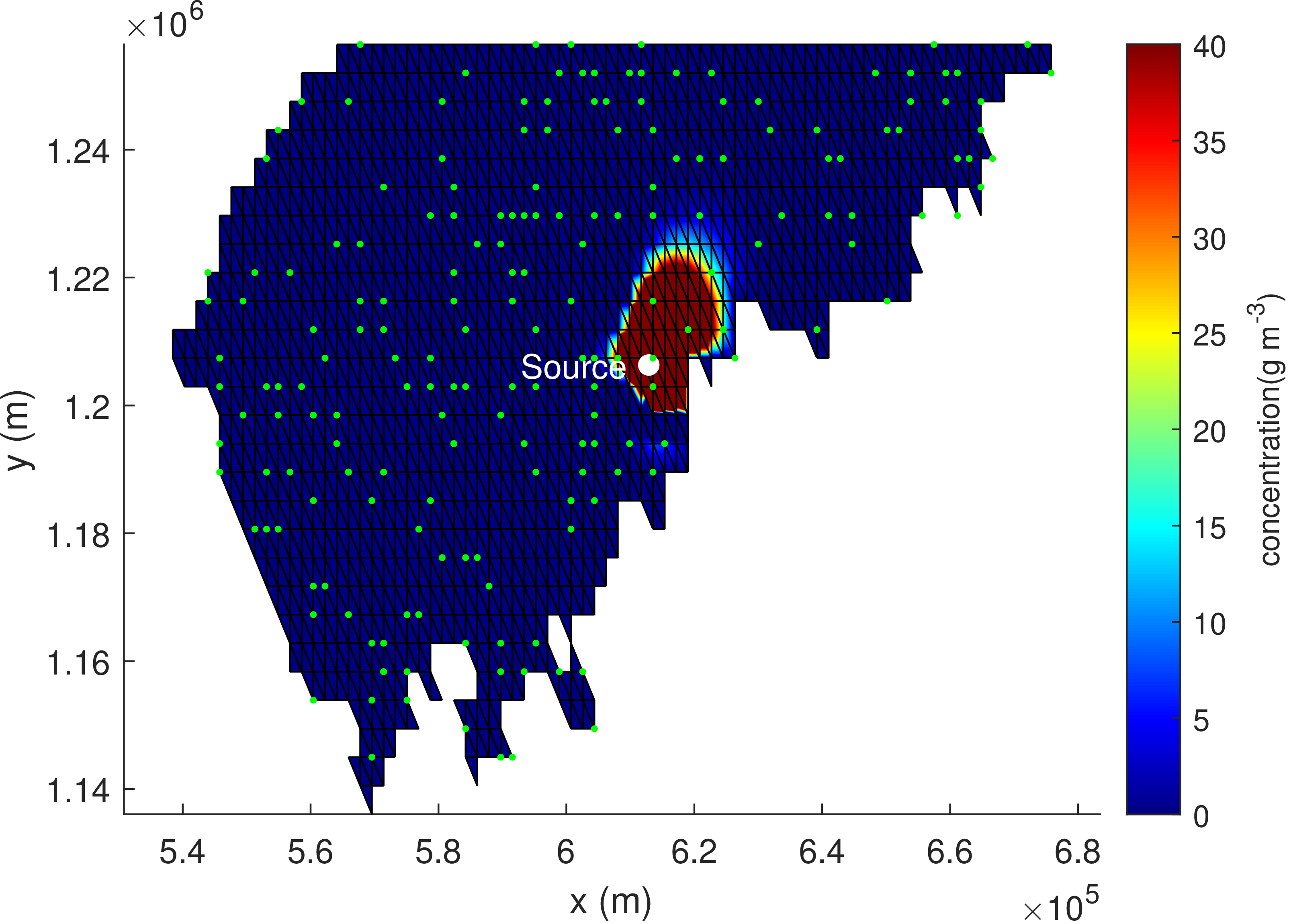}\label{EnKF_estimate2}\end{minipage}
    	}
	\subfigure[$k=18$]{
		\begin{minipage}[b]{0.4\textwidth}
			\includegraphics[width=\textwidth]
			{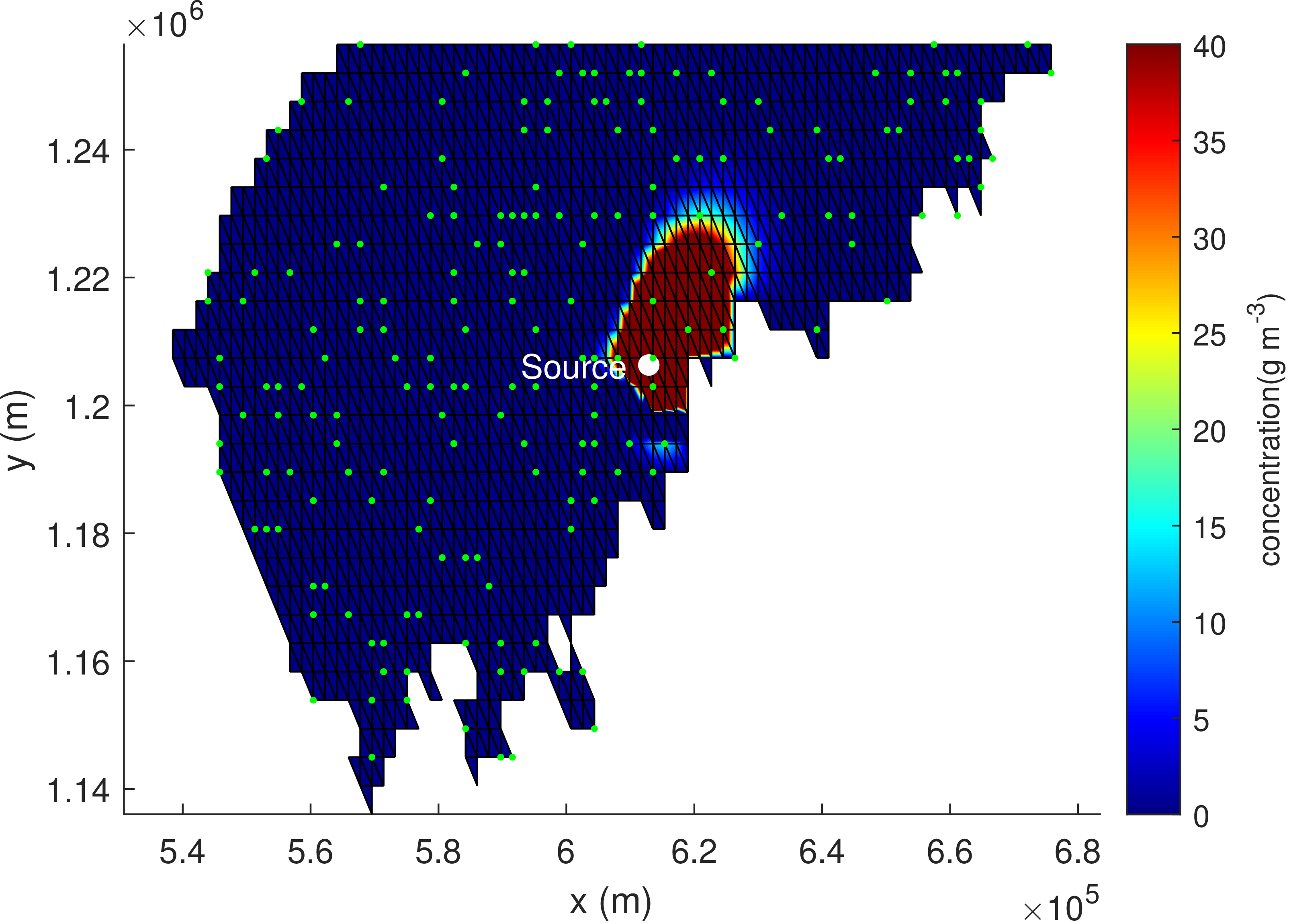}\label{EnKF_estimate3}\end{minipage}
	}
    	\subfigure[$k=24$]{
    		\begin{minipage}[b]{0.4\textwidth}
		 	\includegraphics[width=\textwidth]
		 	{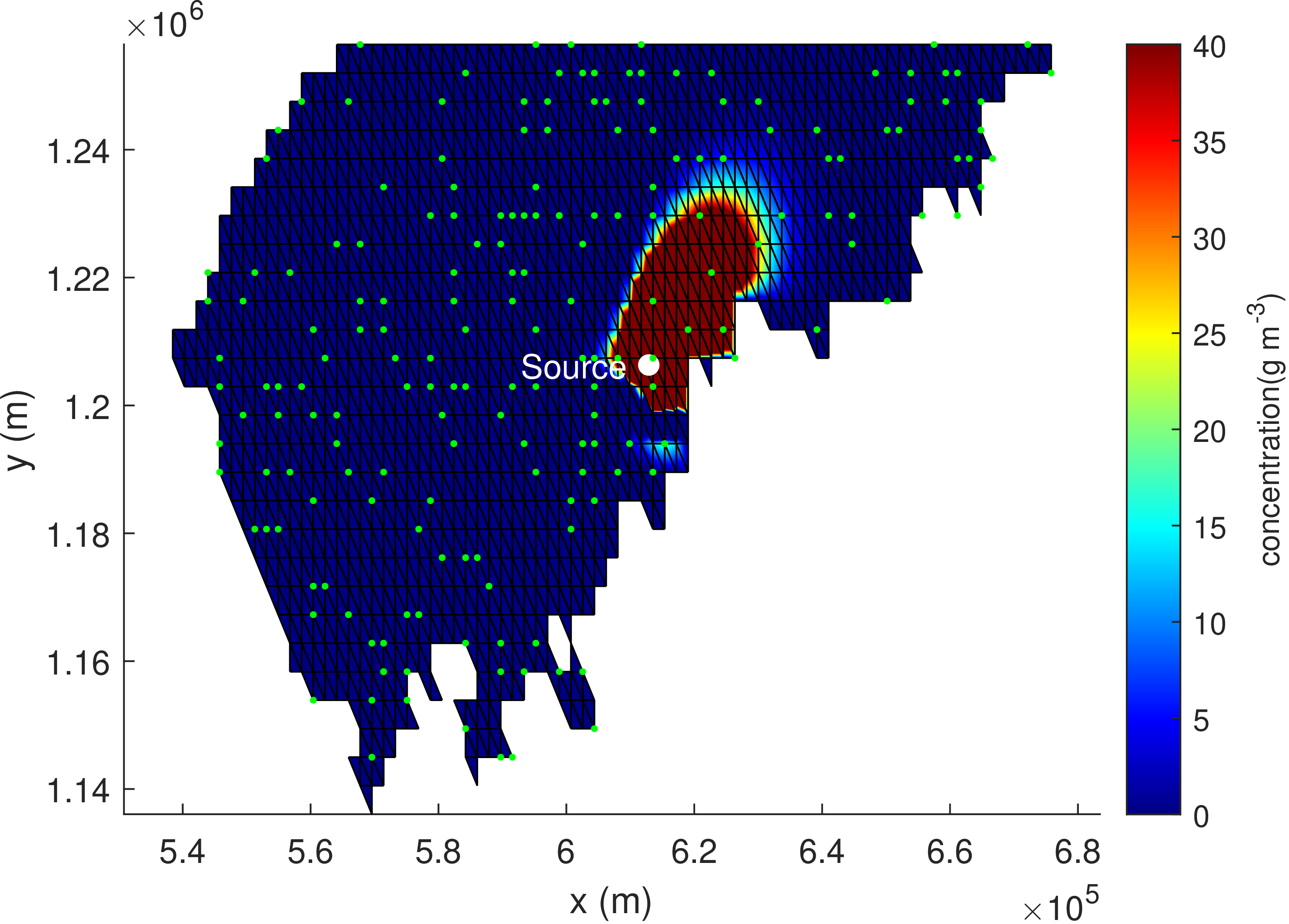}\label{EnKF_estimate4}\end{minipage}
    	}
    		\\ 
	\subfigure[$k=30$]{
		\begin{minipage}[b]{0.4\textwidth}
			\includegraphics[width=\textwidth]
			{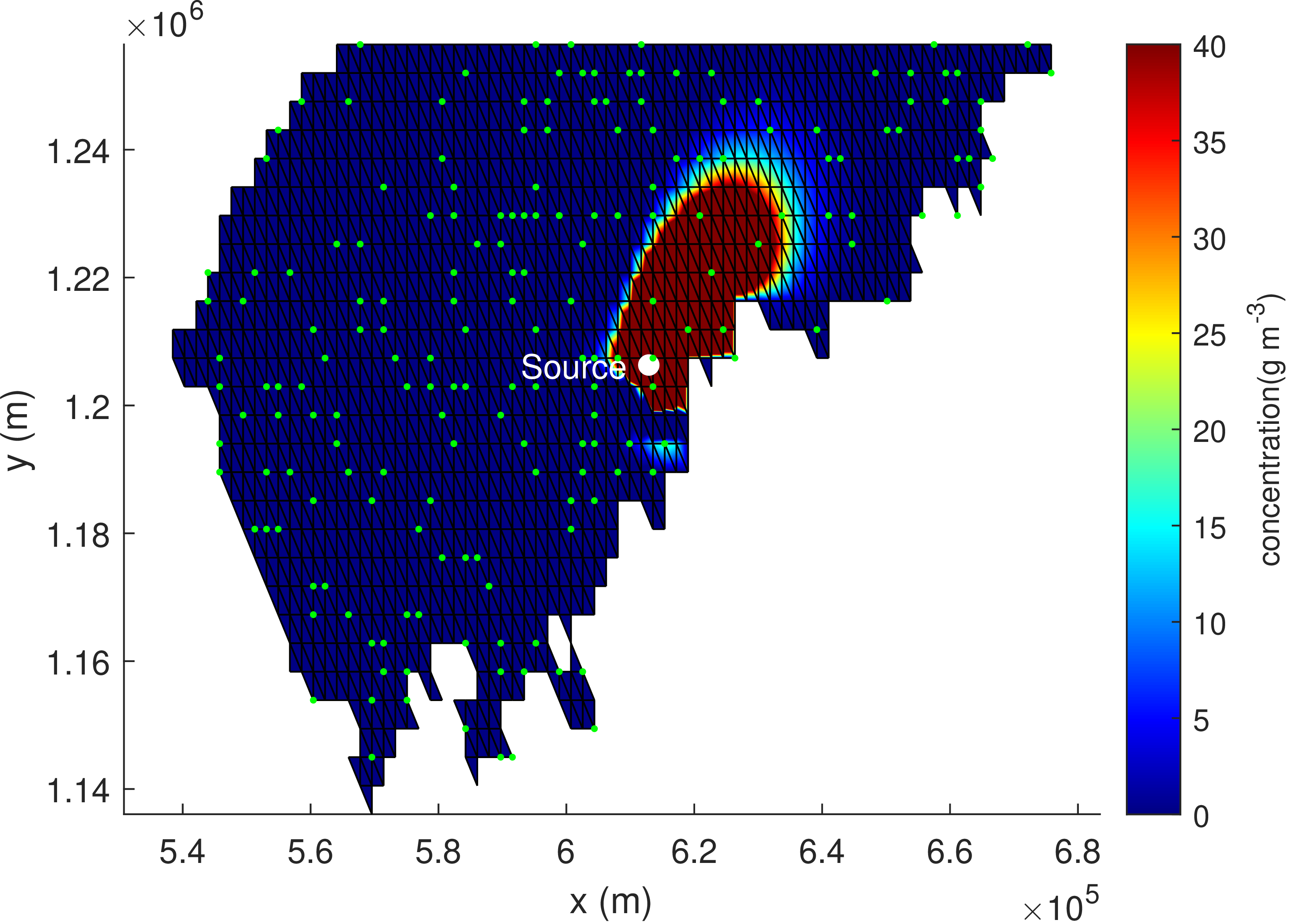}\label{EnKF_estimate5}\end{minipage}
	}
    	\subfigure[$k=36$]{
    		\begin{minipage}[b]{0.4\textwidth}
   		 	\includegraphics[width=\textwidth]
   		 	{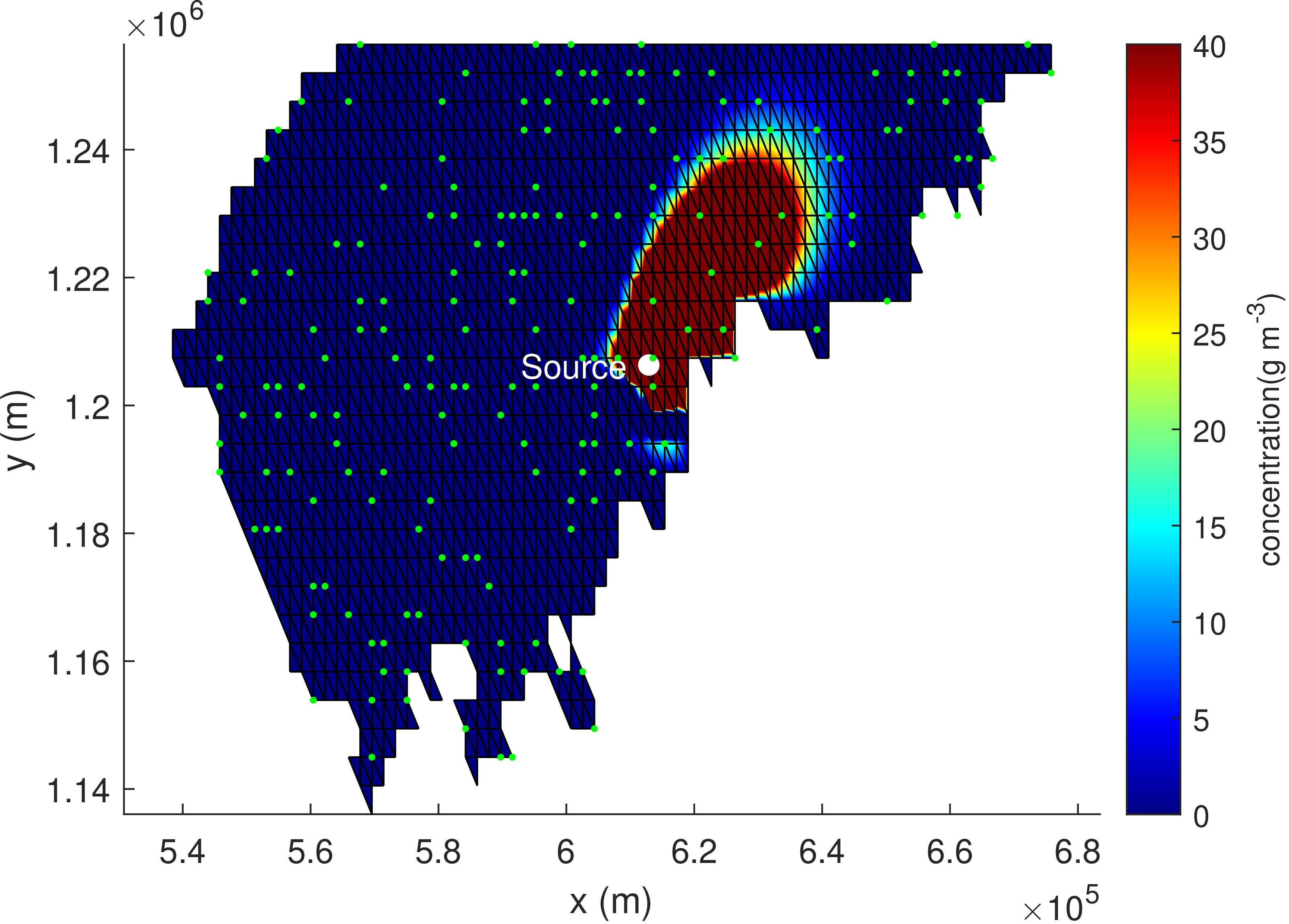}\label{EnKF_estimate6}\end{minipage}
    	}
	\subfigure[$k=42$]{
		\begin{minipage}[b]{0.4\textwidth}
			\includegraphics[width=\textwidth]
			{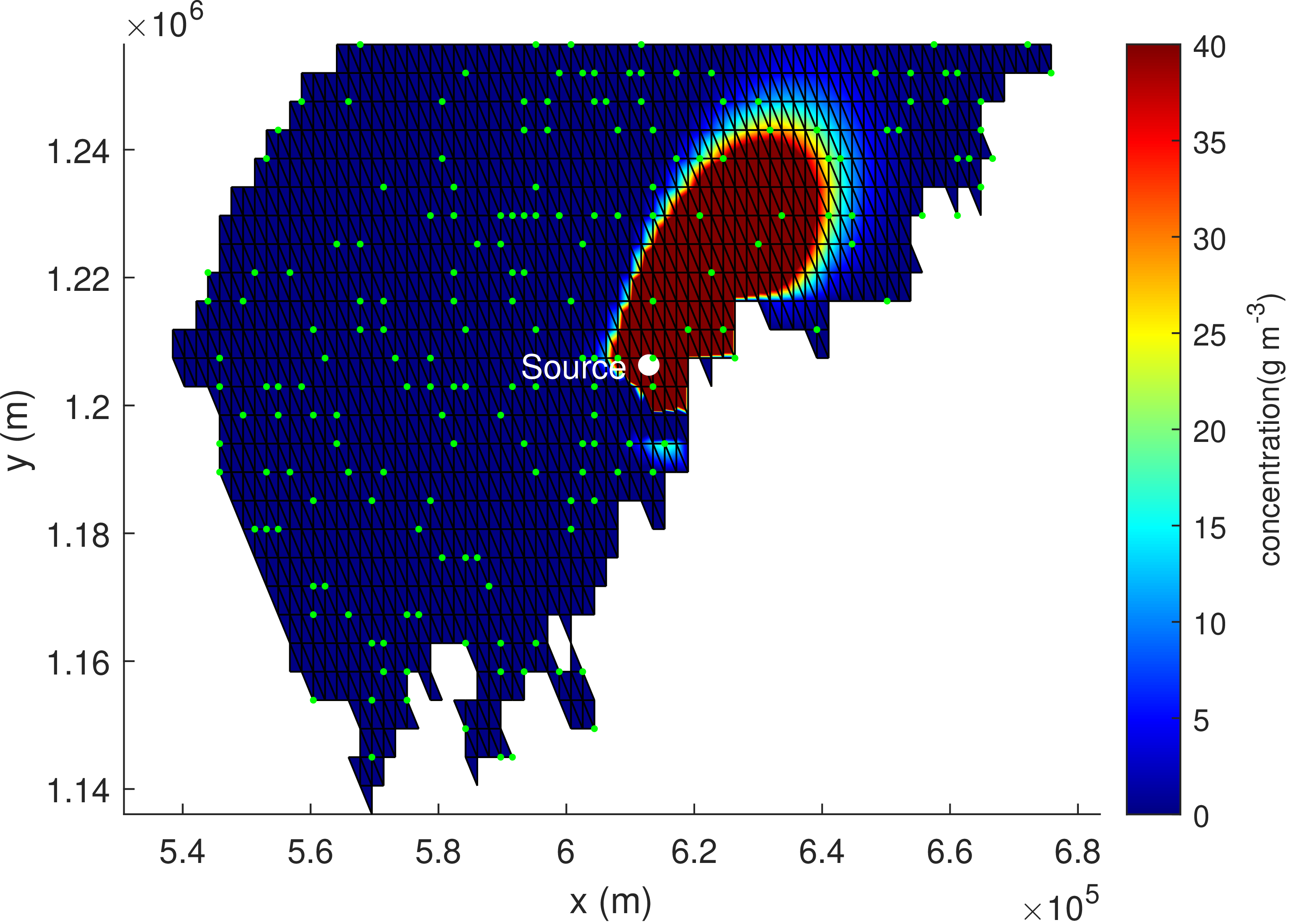}\label{EnKF_estimate7}\end{minipage}
	}
    	\subfigure[$k=48$]{
    		\begin{minipage}[b]{0.4\textwidth}
		 	\includegraphics[width=\textwidth]
		 	{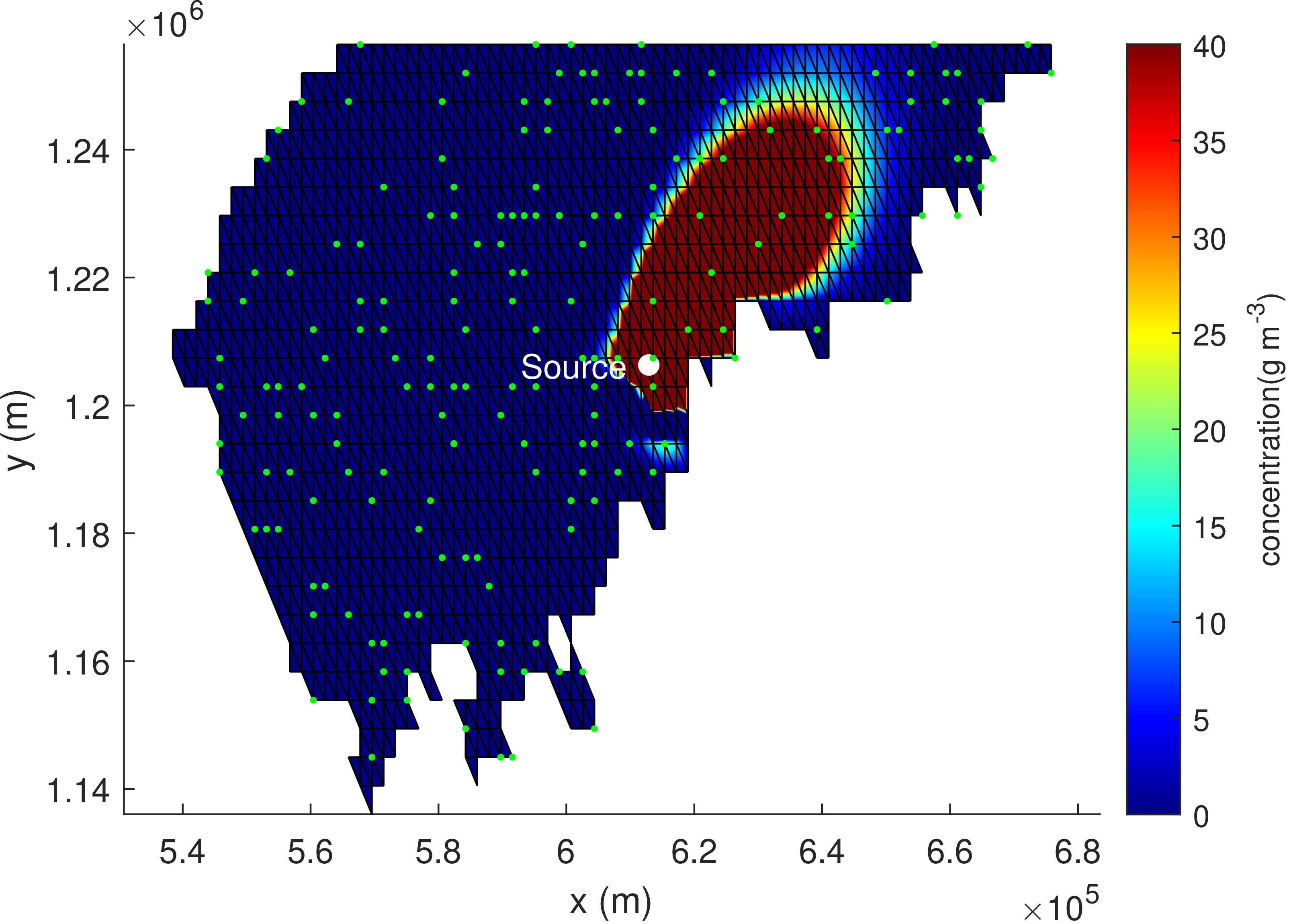}\label{EnKF_estimate8}\end{minipage}
    	}
	\caption{Dispersion process estimates with 30-ensemble EnKF at different sampling instants}
	\label{EnKF_estimate}
\end{figure*}

To investigate the property of the proposed RBPF algorithm, including the effect of the particle number on the estimation results, and to compare the RBPF method with existing approaches, some comparative results are provided in this subsection. EnKF has been used in many environmental researches \citep{cai2022prediction}, including the source estimation for the river pollution \citep{zhang2017ensemble}, which therefore is used as a benchmark algorithm in this paper. The weights of the ensembles in EnKF are equal, which constitutes the main difference from the developed RBPF where the weights of particles are adjusted according to the real-time measurements and the importance resampling mechanism. One illustrative run of the dispersion reconstruction process using EnKF is presented in \ref{EnKF_estimate} (with 30 ensembles).

\begin{figure}[htbp]
\begin{center}
  \includegraphics[width=0.6\linewidth]{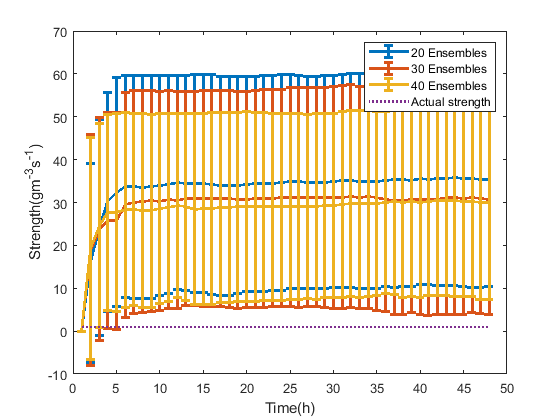}
\caption{Source strength estimates with 30-ensemble EnKF}\label{EnKF_source}
\end{center}
\end{figure}

Setting the ensemble number as the same as the particle numbers used in the previous simulation, the EnKF-based estimation for the source strength over 50 Monte Carlo runs is shown in Figs.~\ref{EnKF_source}. It can be seen that increasing the number of ensembles helps reduce the estimation bias and the standard deviation across different Monte Carlo runs, but the estimation accuracy is much lower than that of RBPF. To quantitatively compare the state estimation performance between the proposed RBPF and the baseline EnKF, the averaged estimation error (AEE) is defined as
\begin{align}\label{AEE}
AEE\triangleq\frac{1}{K}\frac{1}{Q}\sum\limits_{q=1}^{Q}\sum\limits_{k=1}^{K}\|x_k^q-\bar{x}_k^q\|_2,
\end{align}
where $K$ is the total time step, $Q$ is the number of Monte Carlo trials, and $x_k^q$ and $\bar{x}_k^q$ are the realisation of $x_k$ and $\bar{x}_k$ in the $q$th Monte Carlo run, respectively.
The AEE and the execution time obtained with 50 Monte Carlo trials were demonstrated in Table \ref{tbl1} for RBPF and EnKF with different numbers of particles/ensembles. Based on the table, it can be seen that the actual source strength and the dispersion process were estimated well by the developed RBPF. More particles led to slightly more accurate estimates at the cost of more computational time. In comparison, the estimate obtained with EnKF was much less accurate and unduly higher. The estimation results with EnKF were not very consistent, which can be observed from the large standard deviation from the Monte Carlo simulations in Fig.~\ref{EnKF_source}. The performance difference resulted from the adjustment of the particle weights in RBPF. In fact, EnKF can be seen as a group of parallel Kalman filters, and the overall estimation result is the average of the filter outputs with equal weights. In comparison, the weight for each particle in RBPF is determined according to both the system dynamics and the real-time measurements. As such, RBPF outperforms EnKF with same particle numbers at the cost of higher computational load. To draw a fair comparison, we employ a very large number of ensembles, i.e., $350$, in EnKF, so that its computational load is comparable to the proposed RBPF with $20$ particles. The result is added into Table \ref{tbl1}, but its accuracy is still much lower than the proposed RBPF.  


\begin{table*}
\centering
\caption{Averaged Source Strength Estimates with RBPF and EnKF from 50 Monte Carlos runs}\label{tbl1}
\begin{tabular}{l c l l}
\toprule
Method  & Number of ensembles/particles & AEE ($\text{g}\,\text{m}^{-3}\,\text{s}^{-1}$) & Execution time (s)\\ 
\midrule
RBPF & 20 & 8.6107 & 179.1687\\
RBPF & 30 & 8.5949 & 224.7070\\
RBPF & 40 & 8.5264 & 318.2512\\
EnKF & 20 & 37.2645 & 8.5436\\
EnKF & 30 & 39.7313 & 13.3627\\
EnKF & 40 & 31.3172 & 18.4841\\
EnKF & 350 & 22.4243 & 160.2940 \\
\bottomrule
\end{tabular}
\end{table*}

\section{Conclusions} \label{sec4}

An RBPF-based Bayesian estimator was proposed to reconstruct the spatio-temporal distribution of marine surface contaminant and estimate the source emission strength. The general convection-diffusion PDE was solved using the dynamic transient FEM with linear planar triangular elements to approximate the dispersion process from a single source. The FEM model provides a representation of the contaminant dispersion, which is a state-space model; its state values need to be estimated based on the measurement data. Moreover, to account for the potential adversity of imperfect measurements in sensor network-based monitoring, miss detection and signal quantisation was considered in the observation model. A stochastic multiplicative factor was introduced to quantify miss detection. As one of the most-adopted quantisation algorithms, uniform quantisation was selected in our work. 

It is noted that the FEM, discretisation and time integration introduce unavoidable modelling errors, and the consideration of miss detection and signal quantisation introduces nonlinearity and stochasticity. To address these challenges, an RBPF has been adopted to estimate the source strength and dispersion process and both the sensing effects were meticulously resolved in the RBPF design. RBPFs require much fewer particles than conventional PFs, thus more computationally efficient, and have proven to be effective in handling nonlinear and uncertain systems than linearisation based estimation algorithms. 

The effectiveness of the proposed RBPF-based method has been verified based on simulated marine oil spill dispersion in the Baltic sea using real ocean flow data. The results have shown that the developed framework can realise high-performance estimation of spatio-temporal distributions of the leaked crude oil and the source release strength. Additionally, the effects of the particle number have also been analysed, which can help select a proper particle number to achieve a trade-off between computational resources and estimation accuracy. Moreover, a comparative study has been conducted between our proposed approach and EnKF. With the same number of particles/ensembles, the estimation accuracy of our algorithm was much better than that of the EnKF due to the calculation of the particle weights. The Monte Carlo trials have also shown the consistent performance of our approach with small covariance of the estimation results from the trials.

In future work, the two-dimensional model can be extended to the three-dimensional case. Moreover, multiple sources can be included in the same estimation framework. To alleviate the computational burden, reducing the number of states in the model can be considered, for example, by adaptive remeshing or by operating a reduced-order state space for particle filtering. By incorporating these enhancements, the proposed methodology can be further advanced and applied to more complex and realistic scenarios in marine environmental studies, such as the extensively investigated sewage/sludge location problems \citep{schinkel2022synthetic,wang2022automatic,zou2022quinolone}.

\section*{CRediT authorship contribution statement}

Yang Liu: Investigation, Coding, Validation, Visualisation, Writing---Original draft.
Christopher M.~Harvey: Methodology, Coding, Writing---Review and editing, Supervision, Funding acquisition. 
Frederick E.~Hamlyn: Investigation, Coding, Visualisation. 
Cunjia Liu: Conceptualisation, Methodology, Writing---Review and editing, Supervision, Funding acquisition. 

\section*{Data availability}
Data will be made available on request.

\section*{Declaration of competing interest }
The authors declare that they have no known competing financial interests or personal relationships that could have appeared to influence the work reported in this paper.

\section*{Acknowledgement}

This work was supported in part by the NERC Discipline Hopping for Discovery Science 2022 initiative with grant number NE/X018091/1.

\bibliographystyle{apalike}       
\bibliography{ref}  

\end{document}